\documentclass{aa}
\usepackage[varg]{txfonts}
\usepackage{graphicx}

\begin{document}

\title{LMC S154: the first Magellanic symbiotic recurrent nova
  \thanks{Based on observations made with the Southern African Large Telescope (SALT) under programme 2015-2-SCI-036}}


\author{Krystian I\l{}kiewicz\inst{1} \and Joanna Miko\l{}ajewska\inst{1} \and Brent Miszalski\inst{2,3} \and Mariusz Gromadzki\inst{4} \and Berto Monard\inst{5} \and P{\'i}a Amigo\inst{6}} 

\offprints{K. I\l{}kiewicz, \email{ilkiewicz@camk.edu.pl}}

\institute{Nicolaus Copernicus Astronomical Center, Polish Academy of Sciences, Bartycka 18, 00716 Warsaw, Poland
  \and South African Astronomical Observatory, PO Box 9, Observatory, 7935, South Africa
  \and Southern African Large Telescope Foundation, PO Box 9, Observatory, 7935, South Africa
  \and Warsaw University Observatory, Al. Ujazdowskie 4, 00-478 Warszawa, Poland
  \and Kleinkaroo Observatory, Calitzdorp, Western Cape, South Africa 
  \and Instituto de F{\'i}sica, Pontificia Universidad Cat{\'o}lica de Valpara{\'i}so, Casilla 4059, Valpara{\'i}so, Chile
}

\date{Received ... / Accepted ...}

\abstract {  Classical nova outburst has been suggested for a number of extragalactic symbiotic stars, but in none of the systems has it been proven. In this work we study the nature of one of these systems, LMC S154. We gathered archival photometric observations in order to determine the timescales and nature of variability in this system. Additionally we carried out photometric and spectroscopic monitoring of the system and fitted synthetic spectra to the observations. Carbon abundance in the photosphere of the red giant is significantly higher than that derived for the nebula, which confirms pollution of the circumbinary material by the ejecta from nova outburst. The photometric and spectroscopic data show that the system reached quiescence in 2009, which means that for the first time all of the phases of a nova outburst were observed in an extragalactic symbiotic star. The data indicate that most probably there were three outbursts observed in LMC S154, which would make this system a member of a rare class of symbiotic recurrent novae. The recurrent nature of the system is supported by the discovery of coronal lines in the spectra, which are observed only in symbiotic stars with massive white dwarfs and with short-recurrence-time  outbursts. Gathered evidence is sufficient to classify LMC S154 as the first bona fide extragalactic symbiotic nova, which is likely a recurrent nova. It is also the first nova with a carbon-rich donor. }

\maketitle

\section{Introduction}

Symbiotic stars \textbf{(SySts)} are interacting binaries with orbital periods ranging from hundreds of days to decades \citep{2009AcA....59..169G,2013AcA....63..405G}. In these systems matter is transferred from a red giant (RG) to a compact companion, typically a white dwarf (WD). The matter could be transferred through a Roche-lobe overflow or wind accretion. Symbiotic stars are classified as D-type ($dust$) systems when the mass donor is a Mira embedded in a dense nebula. In S-type ($stellar$) systems there is a normal RG as a mass donor (see \citealt{2012BaltA..21....5M} for a recent review). 

Symbiotic novae (SyNe) are SySts in which the WD has experienced a thermonuclear explosion on its surface. They form a small subclass of classical novae and are very rare among SySts. There are two kinds of SyNe. In typical SyNe the outburst occurs on a medium-mass WD and is relatively quiet. In the case of a massive WD accreting at a high rate the outburst is faster and has a higher amplitude. These outbursts are typically observed more then once, hence they are classified as symbiotic recurrent novae (SyRNe). Thus far, only four SyRNe have been identified (T~CrB, V745~Sco, V3890~Sgr and RS~Oph), which makes this class of systems extremely rare.  A recent review of the outburst activity of SySts is presented in \citet{2010arXiv1011.5657M}.

Nine SySts have been discovered in the Large Magellanic Cloud (LMC) and several dozen SySts were discovered outside of the Milky Way in general \citep{2000A&AS..146..407B,2008MNRAS.391L..84G,2009MNRAS.395.1121K,2014MNRAS.444L..11M,2014MNRAS.444..586M,2015AcA....65..139H,2017A&A...606A.110I,2018MNRAS.476.2605I,2018arXiv181106696I}. Outburts of Z~And-type, which are thought to be caused by increased mass transfer onto a WD, have only been observed in the cases of two extragalactic SySts: \object{LIN 9} \citep{2014MNRAS.444L..11M} and \object{LHA 120-S 63}=\object{LMC S63} \citep{2015MNRAS.451.3909I}.  Classical
nova outburst has not been proven for any of the extragalactic SySts.

\object{LMC S154} = \object{LHA 120-S 154} = \object{NSV 16200} was an X--ray source located in the LMC observed during a High-Energy Astronomy Observatory (HEAO)~1 survey \citep{1984ApJS...56..507W} between August 15 1977, and February 15 1978. \citet{1992ApJ...396..668R} failed to detected the object in EXOSITE satellite observations from October 1984. \object{LMC S154} has not been detected in X--rays since that time. \citet{1992ApJ...396..668R} carried out optical observations of the object. The observations carried out during the period from February 2 1984, to February 21 1988, showed that the spectrum of the object was characterized by a flat continuum, strong Balmer-series emission lines, and the presence of some low-excitation emission lines. In observations from December 16 1988, \object{LMC S154} showed a changed spectrum characterized by high-excitation emission lines typical for a SySt. Photometric observations of \citet{1992ApJ...396..668R} showed variability with an amplitude of $\sim$4~mag on a timescale of decades.

The variability of \object{LMC S154} resembles that of SyNe such as for example \object{RX Pup} and \object{PU Vul} \citep{1992ApJ...396..668R,1994A&A...288..842V}. \citet{1994A&A...288..842V} measured overabundance of nitrogen in the nebula around the system, which supports classical nova outburst in the recent history of the system.  \citet{1996A&A...307..516M} detected a C-rich giant in the spectrum of \object{LMC S154} for the first time, which proves the symbiotic nature of the system. The authors derived the spectral type of the cool component to C2,2. \citet{2004RMxAC..20...33M}, on the other hand, suggested that the system was of D-type. The SyN nature of the system has had not yet been confirmed, mainly because of a significant gap in photometric observations published by \citet{1992ApJ...396..668R} ranging from the 1950s to 1980s that hinders estimates of the timescale of its variability. Moreover, the system has not yet been observed in quiescence.

In this work we study \textbf{the possible evidence for} the nova outburst of \object{LMC S154}. We present results of our spectroscopic and photometric survey supplemented with collected archival observations \textbf{in} Sect.~\ref{obs_sec}. The timescale of outburst(s) activity of the system is analyzed in Sect.~\ref{sec:timescale}. The discovery of coronal lines in the system is discussed in Sect. \ref{sec:coronal}. The variability during quiescence is investigated in Sect.~\ref{sec:quiscence}. The physical parameters \textbf{of the cool component} are studied in Sect. \ref{sec:cold}. The evolution of WD parameters during the most recent outburst is presented in Sect. \ref{sec:hot}. Arguments supporting the SyN nature of the system are presented in Sect. \label{sec:novaclass}6.

\section{Observations}\label{obs_sec}

\subsection{Spectroscopy}

\begin{figure*}[h!]
  \resizebox{\hsize}{!}{\includegraphics{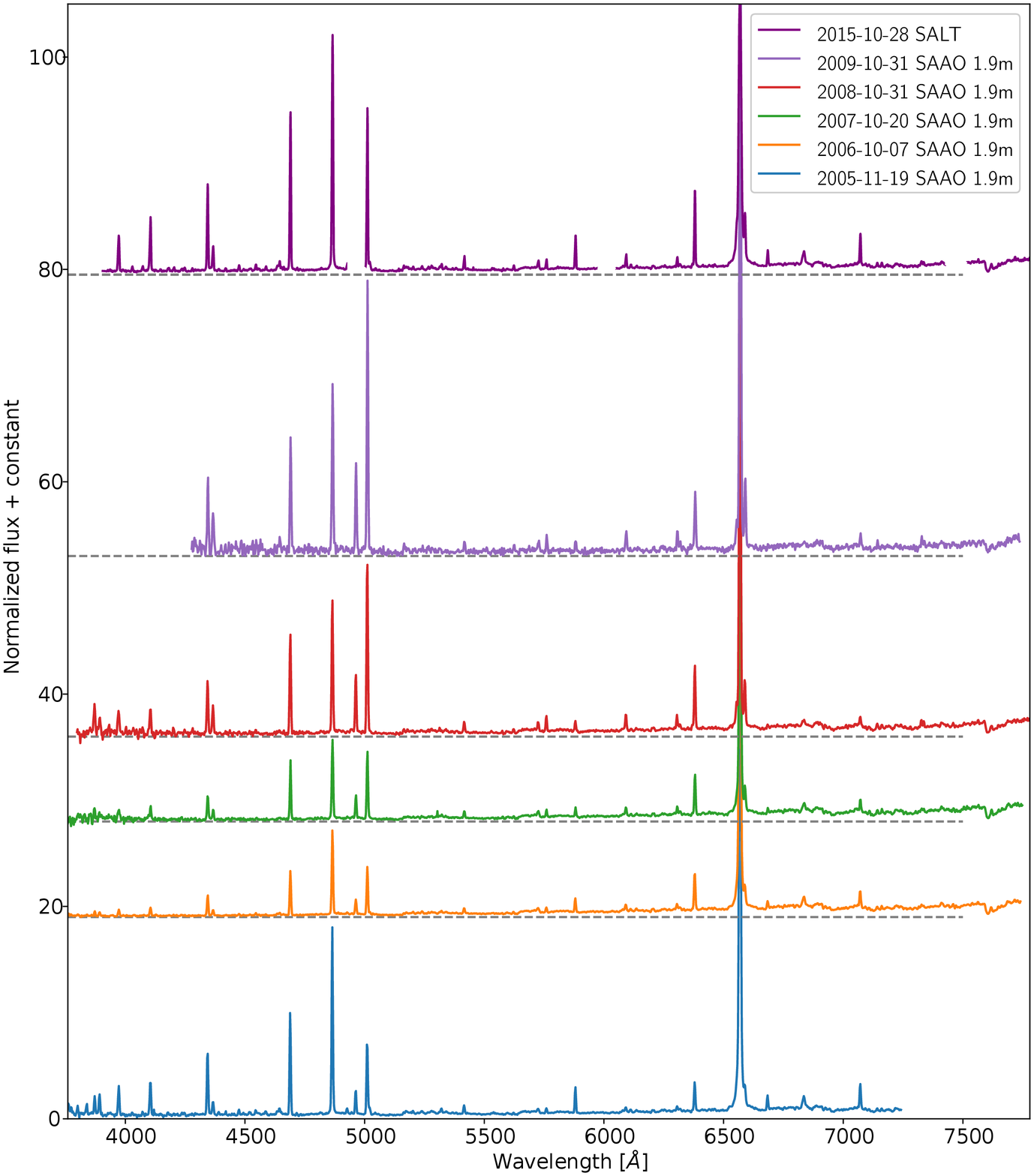}}
  \caption{Spectral variability of \object{LMC S154}. The spectra have been normalized in wavelength range 6700--6800\AA.}
  \label{specfig}
\end{figure*}

We carried out spectroscopic monitoring of \object{LMC S154} using the 1.9-m telescope at the South African Astronomical Observatory (SAAO). The telescope was mounted with the
SITe  (Scientific Imaging  Technologies,  INC.) CCD together with a grating spectrograph. The instrumental setup included grating \#7 with 300 lines per millimeter$^{}$ and a slit width of 1.5''. The resulting resolving power was R$\simeq$1000. The covered spectral range varied with every observation, but approximately covered a range of 3800--7700\AA.

Another low-resolution spectrum was carried out with the  Southern African Large Telescope (SALT; \citealt{2006SPIE.6269E..0AB,2006MNRAS.372..151O}) equipped with the Robert Stobie Spectrograph (RSS; \citealt{2003SPIE.4841.1463B,2003SPIE.4841.1634K}) and PG900 grating. Observations were obtained under programme 2015-2-SCI-036 (PI: Miko\l{}ajewska). \object{LMC S154} was observed in two wavelength ranges giving the combined covered wavelength range of $\sim$3900-8250~\AA. 

The spectra obtained using  SALT and SAAO~1.9-m telescopes were reduced using the standard IRAF procedures. Spectra were flux-calibrated using spectrophotometric standards. Due to slit losses in SAAO 1.9-m data, as well as the variable pupil design of SALT precluding absolute spectrophotometric calibration, we scaled the flux scale of our spectra to the known V-band photometry near the epoch of observation. The list of spectroscopic observations is presented in Table~\ref{table:log}. The spectra are presented in Fig.~\ref{specfig}.

\begin{table}
\caption{The log of spectroscopic observations.}
\label{table:log}
\centering
\begin{tabular}{ccccc}
\hline\hline
Date & MJD      & Telescope     & Exposure [s]  \\
\hline
2005-11-19 & 53693 & SAAO 1.9m & 3x1800  \\
2006-10-07 & 54015 & SAAO 1.9m & 3x1800  \\
2007-10-20 & 54393 & SAAO 1.9m & 3x1200  \\
2008-10-31 & 54770 & SAAO 1.9m & 3x1200  \\
2009-10-31 & 55135 & SAAO 1.9m & 3x1200  \\
2015-10-28 & 57323 & SALT & 2x60,2x1800\\
\hline
\end{tabular}
\end{table}

\begin{figure}
  \resizebox{\hsize}{!}{\includegraphics{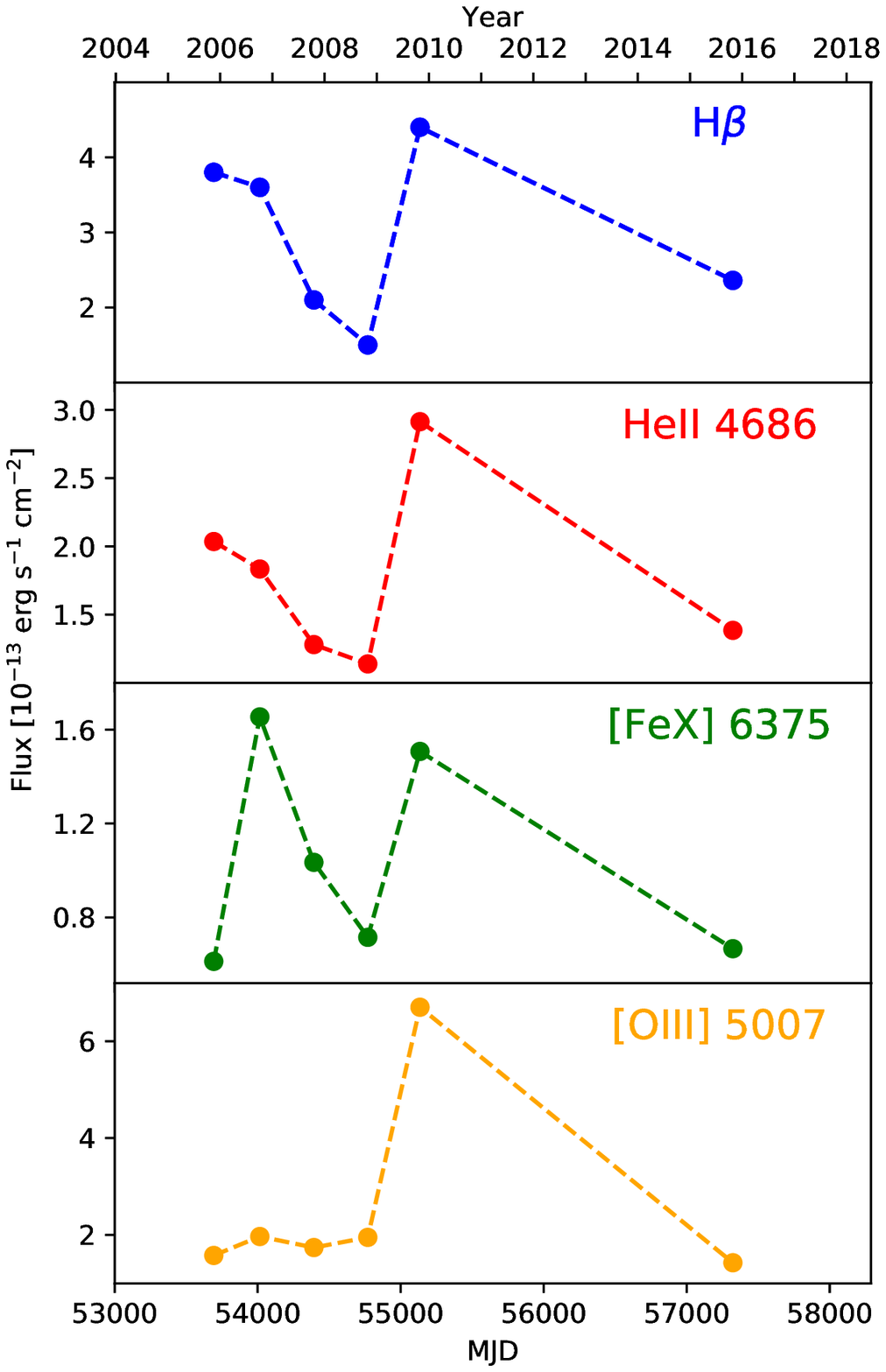}}
  \caption{Variability of emission line fluxes in \object{LMC S154}.}
  \label{S154_fluxes}
\end{figure}

We measured fluxes of emission lines by fitting a Gaussian profile. In the case of lines that could not be fitted by a Gaussian profile we integrated the flux above the local continuum. The main source of uncertainty was choosing the local continuum level. Typical uncertainty was of the order of 15\% in the case of the strongest lines, and 30\% in the case of weakest. The measurements are presented in Table~\ref{table:lineflux} and Fig.~\ref{S154_fluxes}.

\begin{table*}
\caption{Observed wavelength from SALT spectrum and relative fluxes of emission lines.}
\label{table:lineflux}
\centering
\begin{tabular}{cccccccc}
\hline\hline
&Date   &       2005    &       2006    &       2007    &       2008    &       2009    &       2015    \\
&MJD    &       53693   &       54015   &       54393   &       54770   &       55135   &       57323   \\
\hline
ID & $\lambda_{\mathrm{observed}}$ [\AA]&  \multicolumn{6}{c}{100 $\times$ Flux / Flux(H$\beta$) }\\
\hline
{[\ion{Ne}{III}]}~3869  &       3871.04 &       10      &       5       &                &       26      &               &               \\
H$_8$, \ion{He}{I}      &       3892.73 &       12      &       5       &               &       17      &                &               \\
H$\epsilon$     &       3971.61 &       17      &       8       &               &       20      &                &       15      \\
H$\delta$       &       4104.50 &       20      &       10      &       15      &       20      &                &       21      \\
H$\gamma $      &       4343.67 &       35      &       24      &       31      &       40      &       39      &       33      \\
{[\ion{O}{III}]}~4363   &       4366.14 &       8       &       7       &       12      &       21      &       31      &       11      \\
\ion{He}{II}~4686       &       4689.46 &       54      &       51      &       61      &       76      &       66      &       59      \\
H$\beta$        &       4865.15 &       100     &       100     &       100     &       100     &       100     &       100     \\
{[\ion{O}{III}]}~4959   &       4962.98 &       12      &       17      &       26      &       42      &       52      &                \\
{[\ion{O}{III}]}~5007   &       5010.90 &       41      &       55      &       83      &       130     &       152     &       60      \\
\ion{He}{II}~5412       &       5416.11 &       5       &       7       &       8       &       9       &       6       &       5       \\
{[\ion{N}{II}]}~5755    &       5759.26 &               &       6       &       7       &       11      &       9       &       3       \\
\ion{He}{I}~5876        &       5880.54 &       13      &       16      &       10      &       8       &       7       &       11      \\
{[\ion{Fe}{VII}]}~6086  &       6091.67 &               &       8       &       11      &       12      &       12      &       5       \\
{[\ion{O}{I}]}~6300     &       6305.21 &               &               &       6       &       11      &       11      &       3       \\
{[\ion{Fe}{X}]}~6375    &       6379.89 &       16      &       46      &       49      &       48      &       34      &       28      \\
H$\alpha$ + [\ion{N}{II}]~6548  &       6568.66 &       661     &       1012    &       1000    &       616     &       536     &               \\
{[\ion{N}{II}]}~6584    &       6589.26 &       9       &       21      &       29      &       34      &       46      &       31      \\
\ion{He}{I}~6678        &       6683.70 &       8       &       8       &       5       &                &               &       6       \\
\ion{O}{VI}~6825        &       6835.45 &       15      &       19      &       19      &                &               &       10      \\
\ion{He}{I}~7065        &       7070.96 &       15      &       18      &       12      &                &               &       10      \\
{[\ion{Fe}{XI}]}~7892 &  7897.56   &     &     &     &     &     &   7   \\
\hline
 \multicolumn{2}{c}{V$^\star$ [mag]} &  15.60   &       15.95   &       15.90   &       16.40   &       16.60   &       16.35         \\

 \multicolumn{2}{c}{Flux ( H$\beta$ ) [10$^{-13}$ erg s$^{-1}$ cm$^{-2}$]} &       3.8     &       3.6     &       2.1     &       1.5     &       4.4     &       2.36         \\
\hline
\end{tabular}

\raggedright
\textbf{Notes}: $^\star$Assumed magnitude used for scaling of the spectra. \\
\end{table*}

\begin{figure}
  \resizebox{\hsize}{!}{\includegraphics{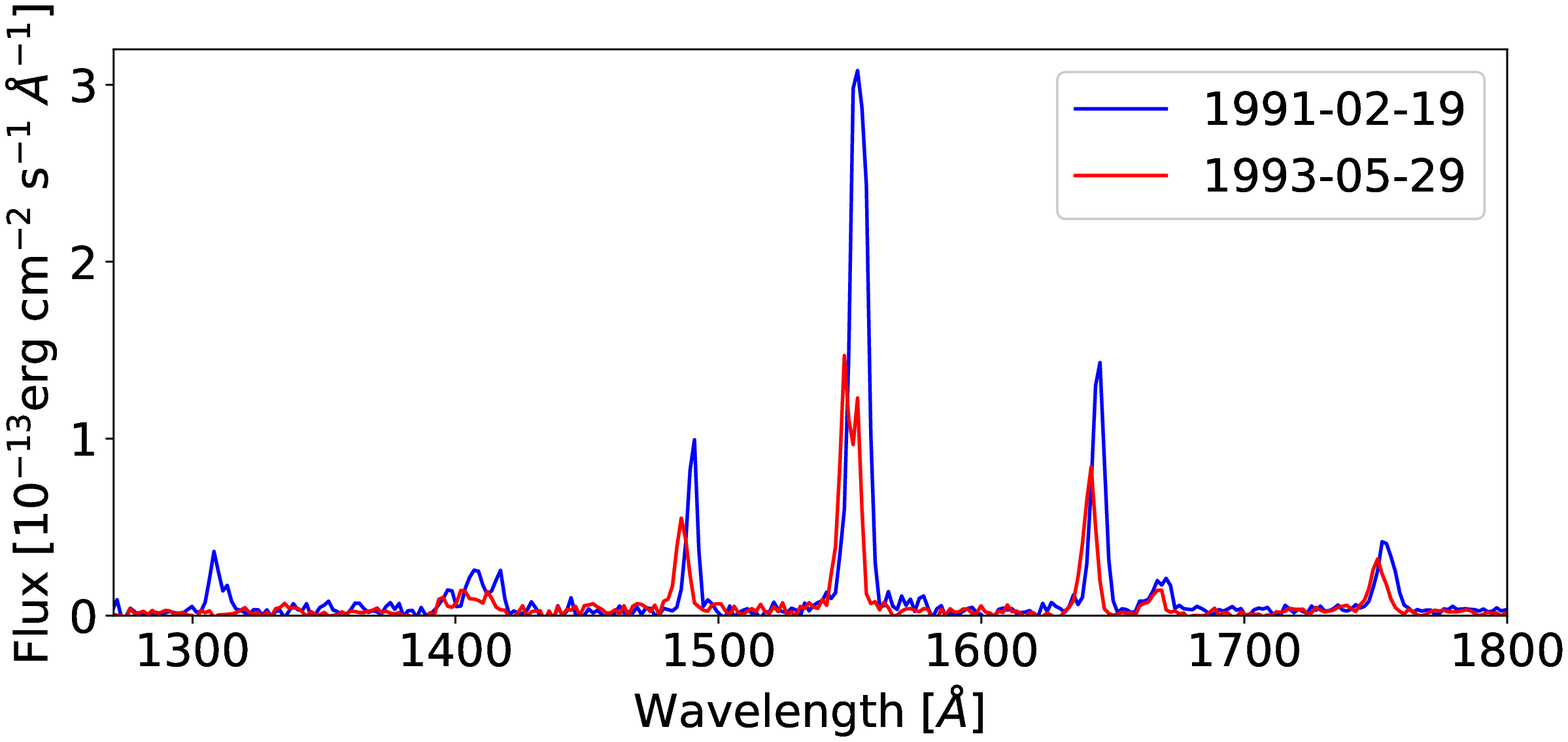}}
  \caption{IUE spectra of \object{LMC S154}.}
  \label{iuefig}
\end{figure}

Additionally we employed two International Ultraviolet Explore (IUE) spectra. The spectra were sp40907 carried out February 2 1991 and sp47766 carried out May 29 1993.  The spectra are presented in Fig.~\ref{iuefig}.

\subsection{Photometry}
Photometric monitoring of \object{LMC S154} was carried out with a 35-cm Meade RCX400 telescope at the Kleinkaroo Observatory. The instrumental setup included the SBIG ST8-XME CCD camera and $V$ and $Ic$ filters. Observations from each night were reduced and stacked in a standard fashion. The photometry was carried out with the single image mode of the AIP4 image-processing software. The accuracy of the derived magnitudes is better than 0.1~mag.

Additionally we observed \object{LMC S154} in $V$ filter with a 2.5-m Irenee du Pont telescope SIT e2K-1 camera at the Las Campanas observatory. Data reduction was performed using standard procedures in IRAF. We employed the brightest stars in the field with magnitudes from the Southern Proper Motion program  (SPM4; \citealt{2011AJ....142...15G}) as standard stars. 

\begin{figure}
  \resizebox{\hsize}{!}{\includegraphics{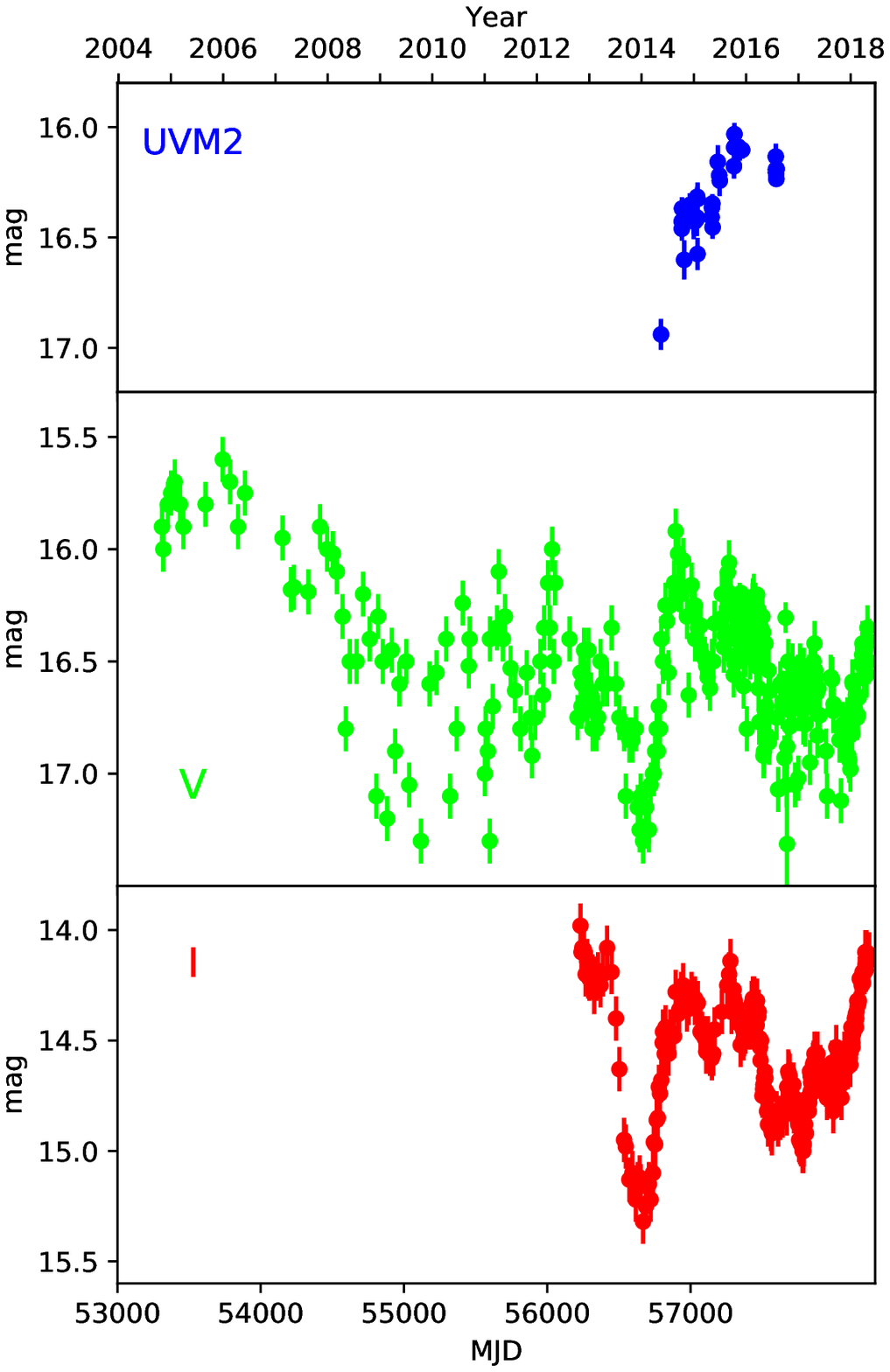}}
  \caption{Most recent photometric variability of \object{LMC S154}.}
  \label{fig:newphot}
\end{figure}

\begin{figure}
  \resizebox{\hsize}{!}{\includegraphics{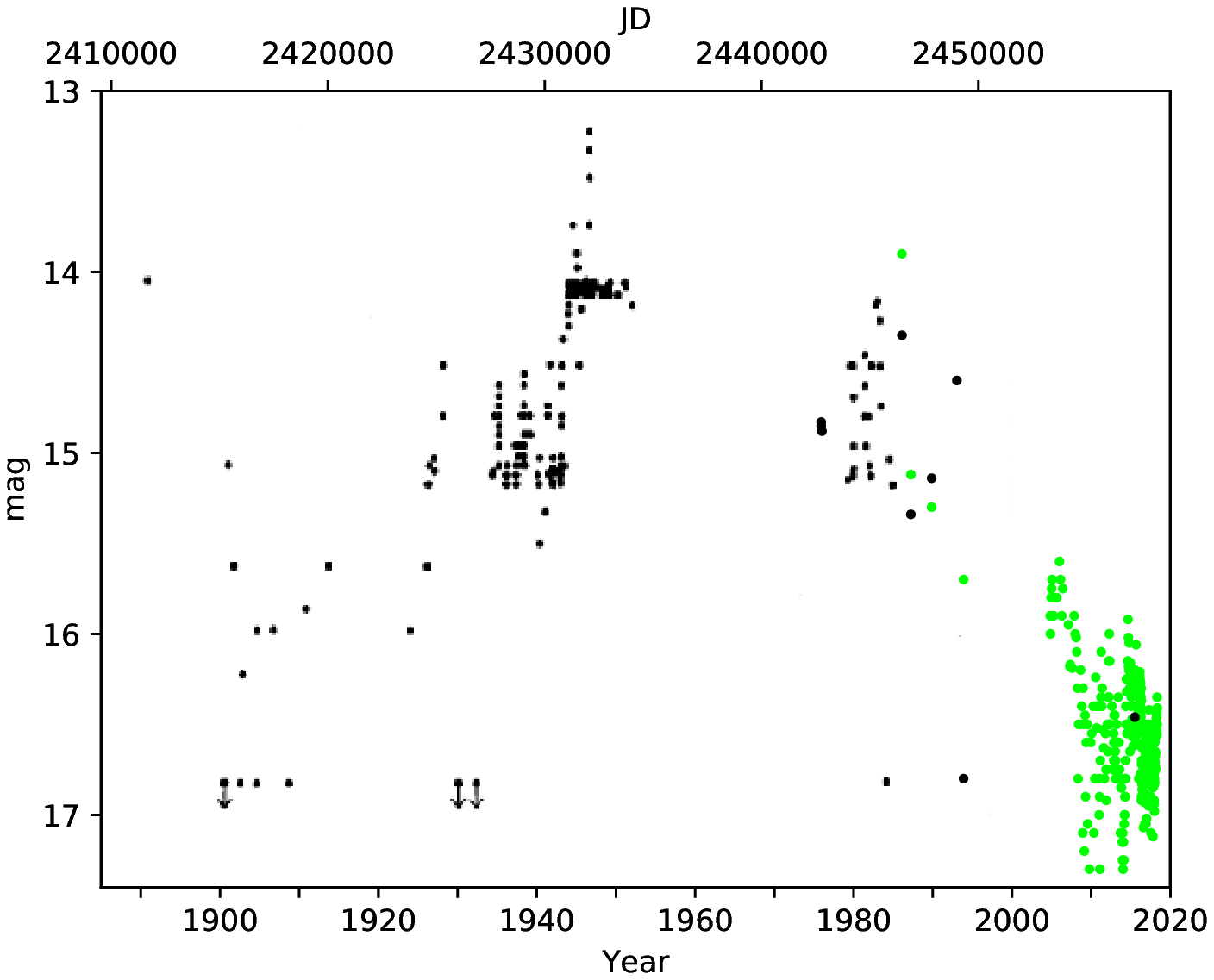}}
  \caption{Historical light curve of \object{LMC S154}. The black points are magnitudes in m$_{\mathrm{pg}}$ and $B$ filters. The green points are magnitudes in $V$ filter.}
  \label{fig:oldphot}
\end{figure}

We supplemented our photometry with collected photometric observations from the literature. The magnitudes are presented in Table~\ref{table:phot}, Fig.~\ref{fig:newphot}, and Fig.~\ref{fig:oldphot}.

We retrieved publicly available UV images containing \object{LMC~S154} taken by the Swift UV optical telescope (UVOT; \citealt{2004ApJ...611.1005G}) from the UK Swift Science Data Centre
(UKSSDC).\footnote{http://www.swift.ac.uk/index.php} Observations were available only in the UV filter UVM2 centered at $\lambda$2246 \AA\ (see \citealt{2008MNRAS.383..627P}). We followed the UVOT data analysis guide of the UKSSDC to perform aperture photometry on the images using the heasoft-6.18 software\footnote{http://heasarc.gsfc.nasa.gov/lheasoft} and the latest caldb files for Swift. Optimal aperture sizes were chosen based on the filter and pixel binning according to \cite{2008MNRAS.383..627P}. Each image was inspected before calculating the magnitudes to ensure the correct object was in the defined aperture. Table~\ref{table:photswift} gives a log of the observations where the magnitudes are provided in the Vega system. The data are presented in Fig.~\ref{fig:newphot}.

\section{Results}
\subsection{Timescale of outburst(s)}\label{sec:timescale}

There is a substantial gap in the photometric observations presented by \citet{1992ApJ...396..668R}. The gap ranges from the 1950s to the 1980s which hinders the determination of whether the \textbf{active phases of \object{LMC S154} in the 1940s and 1980s is the same outburst or a two separate events}. The archival observations presented in Table~\ref{table:phot}  show that in 1975-1976 \object{LMC S154} was at m$_{\mathrm{pg}}\simeq$14.8~mag. This corresponds to the magnitude during an active phase of the system \citep{1992ApJ...396..668R}, hence the last observed active phase started in the 1970s at the latest.

Interestingly, the spectrum of \object{LMC S154} obtained by \citet{1963IrAJ....6...51L} revealed [\ion{O}{III}] lines and no continuum. While \citet{1963IrAJ....6...51L} did not give a date for the observation, most of the data collected for their survey were collected in two seasons in 1956 and 1960 \citep{1975aash.book.....M}, at the beginning of the gap in photometric observations. The presence of [\ion{O}{III}] in the 1950s, and the similarity of spectra observed by \citet{1963IrAJ....6...51L} and \citet{1992ApJ...396..668R}, suggest that \citet{1963IrAJ....6...51L} observed \object{LMC S154} at the nebular phase at the end of outburst, and there were at least two outbursts recorded in the history of \object{LMC S154}, one with a maximum in the 1940s-50s and the second with a maximum in the 1970s-80s. This makes \object{LMC S154} a member of a rare class of SyRNe. Moreover, in the 1890s, despite the scarcity of observations, it seems that the star experienced a drop from m$_{\mathrm{pg}}\simeq$14~mag, a value similar to the one observed at the nova maximum, to m$_{\mathrm{pg}}\simeq$17~mag observed in quiescence (Fig.~\ref{fig:oldphot}). Therefore it is likely that a drop from maximum of an even older nova outburst was observed at the end of the nineteenth century.

In 2009 the system stopped fading and now shows similar $B$ magnitude to that seen in the 1910s (Fig.~\ref{fig:oldphot}). This means that the system finally reached quiescence and the outburst that started in 1970s lasted at least $\sim$40 years. The H$\alpha$/H$\beta$ ratio was gradually decreasing in the observations of \citet{1992ApJ...396..668R},  while \object{LMC S154} was going out of outburst, and the H$\alpha$/H$\beta$ ratio on all of our spectra is in agreement with the latest spectra of \citet{1992ApJ...396..668R}. This supports the idea that \citet{1992ApJ...396..668R} observed the start of the transition to quiescence. The $\sim$40-year timescale of outburst is consistent with approximately two outbursts per century. This also implies a period of $\sim$10-20 years in quiescence after an outburst. This period is consistent with the observed quiescence in the 1900s and 1910s between two outbursts (Fig.~\ref{fig:oldphot}).

\subsection{Coronal lines}\label{sec:coronal}

\begin{figure}
  \resizebox{\hsize}{!}{\includegraphics{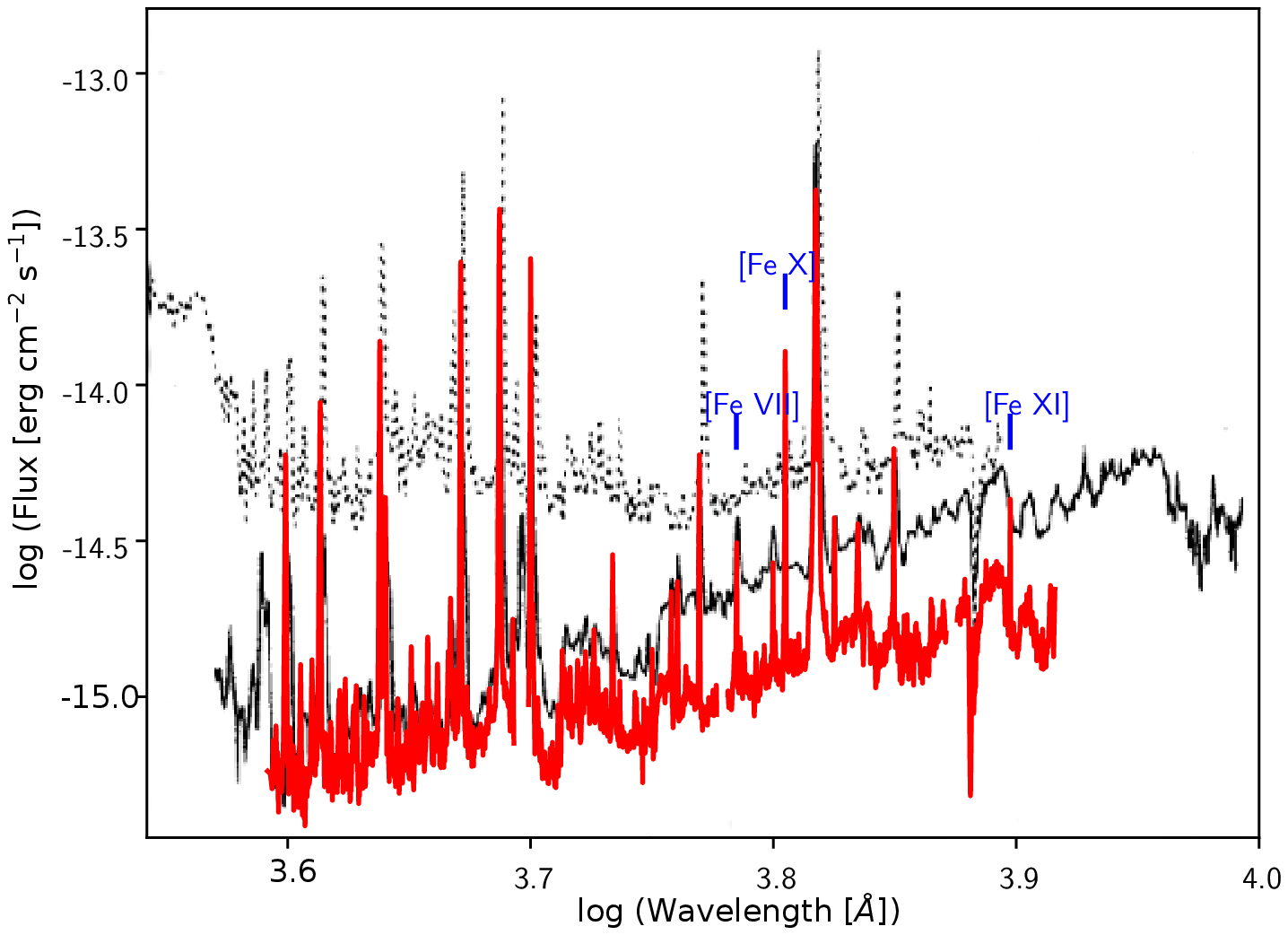}}
  \caption{Comparison of the most recent SALT spectrum of \object{LMC S154} from 2015 (red line) to the spectra obtained by \citet{1992ApJ...396..668R} in 1989 (black dashed line) and \citet{1996A&A...307..516M} in 1993 (black line).}
  \label{fig:refpec}
\end{figure}

The [\ion{Fe}{VII}]~6086, [\ion{Fe}{X}]~6375, and [\ion{Fe}{XI}]~7892 coronal lines are present in our spectra (see Table~\ref{table:lineflux}). [\ion{Fe}{X}] and [\ion{Fe}{XI}] lines are detected for the first time in \object{LMC S154}, while the [\ion{Fe}{VII}]~6086 line  appears to be present in the archival spectra, but was not discussed by authors (see Fig.~\ref{fig:refpec}). The   [\ion{Fe}{X}] and [\ion{Fe}{XI}] coronal lines are very rare in SySt in general (see e.g., \citealt{1996A&A...312..897J} and \citealt{2014MNRAS.444..586M}), but are more common in novae during a nebular phase \citep{1538-3881-144-4-98} and in supersoft X-ray sources (SSXS), where there is stable hydrogen burning on the surface of the WD (e.g., SMC~3; \citealt{1996A&A...312..897J}). Therefore, the presence of [\ion{Fe}{X}] and [\ion{Fe}{XI}] lines in SySt is always associated with nuclear reactions on the surface of WDs.

The [\ion{Fe}{X}]~6375 line is observed during the SSXS phase of a nova outburst, when nuclear burning is still continuing \citep{1985ApJ...294..263M,1989ApJ...341..968K}.  Hence, the [\ion{Fe}{X}]~6375 line is well correlated with soft X-ray flux, both in SyNe \citep{2012A&A...540A..55S} and in classical novae \citep{2001AJ....121.1636M}. Therefore, it is expected that in \object{LMC S154} the X-ray emission should be the strongest in SSXS phase, around the maximum of [\ion{Fe}{X}]~6375 in the 2000s. However, the X-ray emission was detected in the 1970s, and it was not present in the 1980s \citep{1992ApJ...396..668R}. Since a WD can experience only one SSXS phase during
an outburst, this indicates that the  X-ray emission in the 1970s was not due to the SSXS phase, but originated in a shock produced by an expanding blast wave \citep{2006Natur.442..276S,2009ApJ...691..418D}. This mechanism of radiation is expected at the earliest phase of the nova outburst  is therefore a strong argument in support of the idea that the most recent outburst started in the 1970s.

Assuming a distance to LMC of 49.97~kpc \citep{2013Natur.495...76P} a reddening E($B-V$)=0.17~mag \citep{1996A&A...307..516M} and using [\ion{Fe}{X}]~6375 flux from the 2015 SALT spectrum  we get  L([\ion{Fe}{X}]~6375)=2.0$\times$10$^{34}$~erg~s$^{-1}$. Maximum  [\ion{Fe}{X}]~6375 flux observed during the last outburst of RS~Oph  was 1.5$\times$10$^{-10}$~erg~cm$^{-2}$~s$^{-1}$ \citep{2009A&A...505..287I}. Assuming a distance to RS~Oph of 1.4~kpc \citep{2008ASPC..401...52B} and reddening E($B-V$)=0.69~mag \citep{2018MNRAS.480.1363Z} this corresponds to a luminosity L([\ion{Fe}{X}]~6375)=1.9$\times$10$^{35}$~erg~s$^{-1}$. In SMC~3, an SSXS SySt, the [\ion{Fe}{X}]~6375 line was 2.0$\times$10$^{-14}$~erg~cm$^{-2}$~s$^{-1}$ \citep{1992MNRAS.258..639M} which after scaling for reddening E($B-V$)=0.099~mag \citep{2005MNRAS.357..304H} and SMC distance of 60.6~kpc \citep{2005MNRAS.357..304H}  gives L([\ion{Fe}{X}]~6375)=1.1$\times$10$^{34}$~erg~s$^{-1}$. Given the differences in metallicity and reddening of these objects, one could say that the luminosity of the [\ion{Fe}{X}] line in the two X--ray-bright objects is consistent with   [\ion{Fe}{X}] luminosity in  \object{LMC S154}. This confirms that [\ion{Fe}{X}]  and [\ion{Fe}{XI}] emission lines are due to the SSXS phase of outburst.

Given the spectroscopic evolution of \object{LMC S154} in observations of \citet{1992ApJ...396..668R} the maximum of the nova outburst was most probably no later than their first spectroscopic observation in February~1984. The maximum was probably even earlier given the fact that the star was detected in X-rays in the 1970s. The coronal lines were not present at least until May~1993 \citep{1996A&A...307..516M} which corresponds to at least 9 years after the maximum. The first spectrum with  the [\ion{Fe}{X}]~6375 line detected was from 2005, at least 21~years after the maximum. The emission line was still present more than 31~years after the maximum. 

Detection of this line so late after outburst is somewhat surprising. Typically in the recurrent novae and SyRNe the [\ion{Fe}{X}]~6375 line is detected $\sim$1 month after outburst \citep[see e.g.,][]{1991ApJ...376..721W,2009A&A...505..287I} while in classical novae this line can be detected a few years after the maximum \citep{1989ApJ...341..968K}. Moreover, in SyNe the [\ion{Fe}{X}]~6375 line is observed only in the case of massive WDs that exhibit nova outburst on short timescales; for example in \object{RS Oph} \citep{2011ApJ...727..124O}, \object{V407 Cyg} \citep{2012A&A...540A..55S}, and \object{T CrB} \citep{1953AnAp...16...73B}. In SyNe, with long timescales and less massive WDs, only [\ion{Fe}{VII}] lines are observed; for example in \object{RX Pup} \citep{1999MNRAS.305..190M}, \object{AG Peg} \citep{2001AJ....122..349K}, and \object{PU Vul} \citep{2007JKAS...40...39Y}.  This suggests that the outburst timescale in \object{LMC S154} was in intermediate range ($\ll$100yrs) which is consistent with the hypothesis that there were two separate outbursts, one with its maxumim in the 1940s-50s and the second with its maximum in the 1970s-80s.

\subsection{X-ray and UV emission during outburst}

X-ray radiation was detected in \object{LMC S154} between August 15 1977 and February 2 1978 and has not been detected since \citep{1992ApJ...396..668R}. If there was in fact a nova outburst in the system that came to a halt in the 1990s, the subsequent lack of X-ray radiation is expected. In order to confirm the reality of X-ray radiation in \object{LMC S154} during the outburst we search the High Energy Astrophysics Science Archive Research Center\footnote{http://heasarc.nasa.gov/} (HEASARC)  for additional reported observations. The only finding is the reported detection of \object{LMC S154} by the Eighth Orbiting Solar Observatory (OSO-8) on October 14 1975. The source was detected at 4.7$\pm$0.1~counts~s$^{-1}$ in the range of 2-60~keV. This observation was before the one reported by \citep{1992ApJ...396..668R} and supports the reality of X-ray emission from \object{LMC S154} during the most recent outburst.

Change of emission line fluxes in the two IUE spectra is evident (Fig.~\ref{iuefig}). Fluxes of all of detected emission lines decreased between 1991 and 1993. This is consistent with the drop in brightness of the system during 1993 (Table~\ref{table:phot}). In the spectrum from 1993 the emission lines appear to be narrower. This lead to the separation of \ion{C}{IV}~1548 and \ion{C}{IV}~1551 lines. A similar drop in the width of emission lines after a nova outburst was observed for example in \object{AG Peg} \citep{1993AJ....106.1573K}. This is probably related to the fact that these lines are formed in the wind of the WD, which has slowed down after the outburst. Interestingly there is an apparent blueshift of the emission lines in the spectrum from 1993.

\subsection{Photometric and spectroscopic variability in quiescence}\label{sec:quiscence}

The outburst of  \object{LMC S154} ended in 2009  and the star stopped fading. After the end of the outburst, during 2013 and 2014, \object{LMC S154} showed a dip in its light curve, where the brightness decreased by $\sim$1~mag in $V$ and $I$ bands (Fig.~\ref{fig:newphot}). After recovery from the dip the Swift light curve showed that the star brightened by $\sim$1~mag in 1.5~yr (Fig.~\ref{fig:newphot}). Similar behavior was observed in $V$ band in \object{RX Pup}, where both the dip at the end of the outburst and a subsequent brightening were observed \citep{1999MNRAS.305..190M}. This was interpreted as a signature of ceasing of the hot component wind which allowed for a start of accretion of the cool component wind \citep{1999MNRAS.305..190M,2002AIPC..637...42M}. Another explanation of brightening in the Swift UV observations could be active phases observed in other SyRNe (see e.g., \citealt{2016MNRAS.462.2695I}).

In the most recent observations in $V$ and $I$ bands \object{LMC S154} shows quasi-periodic variability on a timescale of $\sim$250d (Fig.~\ref{fig:newphot}). This variability is not present in the Swift $UVM2$ band. The simplest interpretation would be pulsations of the RG. This is consistent with the fact that the system shows variability with an amplitude of $\sim$0.5~mag in $J$ and $K$ filters (see Table~\ref{table:phot}). Moreover, pulsations would be consistent with the position of the RG on the $K$ versus ($J-K$) colour--magnitude diagram for LMC \citep{2007AcA....57..201S}. 

In the visual spectrum the [\ion{N}{II}] emission lines are detected for the first time. They are clearly visible in all of our spectra (Fig.~\ref{specfig}). The presence of these lines is typical for a nebular phase after a nova outburst (see e.g., \citealt{2008clno.book.....B}). Both the [\ion{N}{II}] and [\ion{O}{III}] nebular lines were roughly constant in the years 2005--2008 (see Fig.~\ref{S154_fluxes}, Table~\ref{table:lineflux}). The fluxes were significantly higher in 2009, which was probably related to the end of the outburst. In the spectrum that followed, carried out in 2015, the forbidden lines had the same flux as during the outburst.

The \ion{H}{} and \ion{He}{} permitted lines showed a gradual decrease in the years 2005--2008 (Fig.~\ref{S154_fluxes}, Table~\ref{table:lineflux}) similar to the one observed in photometric observations (Fig.~\ref{fig:newphot}). In 2009, when the photometric decline had come to a halt, the permitted emission lines showed an increase in flux similar to the one observed in the case of forbidden lines. Moreover, in 2015, both the permitted and emission lines showed similar fluxes to those found at the end of the photometric decline.

In the case of the [\ion{Fe}{X}]~6375 emission line the variability is the most complicated. The behavior of this line is similar to the  variability seen for \ion{H}{} and \ion{He}{} permitted lines. The only difference is that the first two of our spectra show an increase in flux which is succeeded by a gradual decrease. This may be related to stopping of the WD wind, which lead to a decrease in the density of the material close to the WD, and allowed for emission of forbidden lines in a high-excitation environment.

\section{Cool component}\label{sec:cold}

The cool component of \object{LMC S154} was classified as a C-rich giant by \citet{1996A&A...307..516M}. The authors determined its type as C2,2. Nevertheless, thus far the chemical composition of the RG has not been estimated and only the chemical composition of the nebula around the system has been calculated using emission lines \citep{1994A&A...288..842V}. This gave a consistent overabundance of carbon (C/O=1.19). Nevertheless, the outburst in the history of the system can pollute the nebula with material from the WD and cause the abundances to misrepresent the giant chemical composition \citep[see e.g.,][]{2015MNRAS.451.3909I}. For this reason we attempted to estimate the carbon abundance in the cool component of \object{LMC S154} by fitting a synthetic spectrum to the observations.

In order to obtain an initial guess for the RG parameters we fitted  a GRAMS C-rich grid of models of dust shells around red supergiant and AGB stars \citep{2011A&A...532A..54S} to the infrared spectral energy distribution (SED) of \object{LMC S154}. We employed photometry from 2MASS All-Sky Point Source Catalog \citep{2006AJ....131.1163S}, AKARI/IRC Mid-Infrared All-Sky Survey \citep{2010A&A...514A...1I}, and the IRAS Catalog of Point Sources \citep{1988iras....7.....H}. Bayesian analysis of data was performed using the virtual observatory SED analyzer (VOSA) service \citep{2008A&A...492..277B}. We assumed E$_{B-V}$=0.17 \citep{1996A&A...307..516M} and a distance to LMC of d=49.97~kpc \citep{2013Natur.495...76P}. The caveat is that the models were only for a solar metallicity and in the grid of models there was a highest possible $\log g=0$. The results of our analysis are presented in Table.~\ref{table:fitIR}.

\begin{table}
\caption{Parameters of the red giant from analysis of the infrared SED of \object{LMC S154}.}
\label{table:fitIR}
\centering
\begin{tabular}{cccccc}
\hline
 Parameters &  Best value & Probability [\%]\\\hline
 $T_{\rm eff}$ &  4000 & 79.76  \\
L &  12300 L$_\odot$ & 78.29\\
$\log \mathrm{g}$ & 0 & 100   \\
Mass &  2       M$_\odot$ &100  \\
C/O &  2 & 43.15  \\
$\dot{ \mathrm{M}}_{dust}$  & $1.27\times 10^{-9}$ M$_\odot$/yr & 35.75 \\
 $\mathrm{R_{in}}$& 4.5 R$_\mathrm{star}$ & 100  \\
$\tau_{11.3}$ & 0.2 & 100 \\
\hline
\end{tabular}
\tablefoot{
$\dot{M}_{dust}$ -- dust mass-loss rate, $R_{in}$ -- inner radius of the dust shell, $\tau_{11.3}$ -- optical depth at 11.3 microns 
}
\end{table}

\begin{figure}
  \resizebox{\hsize}{!}{\includegraphics{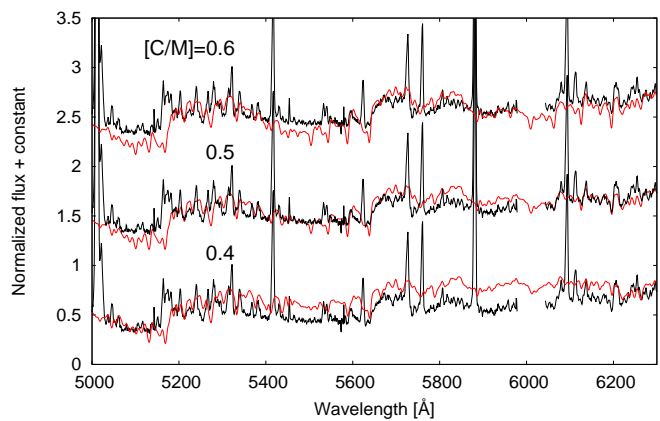}}
  \caption{Fit of synthetic spectra (red lines) to the SALT spectrum of \object{LMC S154} (black lines).}
  \label{fitfig}
\end{figure}

We calculated synthetic spectra using the SPECTRUM code \citep{1994AJ....107..742G} and models from \citet{2004astro.ph..5087C}. During calculations we employed the updated \citet{1998A&A...337..495P} TiO line list from the their website\footnote{http://www.pages-perso-bertrand-plez.univ-montp2.fr/} and solar abundances from \citet{1998SSRv...85..161G}. We assumed LMC metallicity [M/H$]=-0.5$~dex and $\log g=0$. Our grid of synthetic spectra consisted of models with $T_{\rm eff}=3500$--$4500$~K and [C/M$]=0.0$--$1.0$~dex. The results of least square fitting of synthetic spectra to the SALT spectrum of \object{LMC S154} are presented in Fig.~\ref{fitfig}.  Using this procedure we derived $T_{\rm eff}=4000\pm250$ and [C/M$]=0.50\pm0.05$~dex. Additional uncertainty could be introduced by the presence of a nebular contribution to the continuum, which we did not consider.

After assuming the same oxygen abundance as assumed metallicity from our fit we derive C/O$\simeq$1.7. This value is clearly higher then C/O$=$1.19 derived by \citet{1994A&A...288..842V} for the nebula, even taking into account the relatively low accuracy of the \citet{1994A&A...288..842V} method. This confirms that the nebula has been polluted by the material ejected from the WD during the nova outburst.

\section{WD evolution during outburst}\label{sec:hot}

During a maximum of both a SyNe and Z-And outburst the WD experiences a ````supergiant phase'', where the WD resembles an A--F supergiant. During this phase most of the light is shifted towards optical bands. Therefore, we can estimate a luminosity of the WD during maximum by measuring the brightness of the system in the optical bands, assuming that the bolometric correction BC$\simeq$0. The maximum m$_{\mathrm{pg}}$ observed in the archival outburst was $\sim$13.1~mag (Fig.~\ref{fig:oldphot}). Due to gaps in photometric data we assume this is the maximum magnitude during both outbursts. A lower limit of the maximum magnitude of the 1980s outburst is $\sim$14~mag. After correcting for a reddening  E($B-V$)=0.17~mag \citep{1996A&A...307..516M} and using a distance to LMC of 49.97~kpc \citep{2013Natur.495...76P} this method gives a L$_{\mathrm{WD}}\sim$8300--23000L$_\odot$ during the outburst maximum.

At the beginning of the nebular phase, the luminosity of the WD has been estimated to be L$_{\mathrm{WD}}=3000\pm1000$L$_\odot$ in May~1993 \citep{1994A&A...288..842V,1996A&A...307..516M} based on UV spectroscopy. Using our data we can estimate WD luminosity with Eq.~8 of \citet{1991AJ....101..637K}. Fluxes measured in 2015 (Table~\ref{table:lineflux}) after correcting for reddening give L$_{\mathrm{WD}}\sim$4200L$_\odot$. Similarly, the \ion{O}{VI}~6825 flux from the same spectrum gives L$_{\mathrm{WD}}\sim$4700L$_\odot$ using Eq.~8 of \citet{1997A&A...327..191M}. Both of these estimates have an uncertainty of a factor of two.

While there is no data covering the maximum of the outburst allowing us to estimate WD temperature, \citet{1994A&A...282..586M} estimated T$_{\mathrm{hot}}=140\pm15$ based on spectra from 1993. Using our data we can estimate the temperature of the WD using a relation proposed by \citet{1994A&A...282..586M}:
\begin{equation}
\mathrm{T}_{\mathrm{WD}}(\mathrm{1kK})=\chi_{\mathrm{max}}(\mathrm{eV})
,\end{equation}
where $\chi_{\mathrm{max}}$ is the highest observed ionization potential. The highest ionization stage observed in  \object{LMC S154} is Fe$^{+10}$, which gives T$_{\mathrm{WD}}$=291~kK. 

Evolution of the WD during outburst is presented in Table~\ref{table:WDourburst}. During the decline from maximum, the temperature of the WD increased by over a factor two. Such a behavior is expected during decline from classical nova outburst, and was observed for example in the SyN \object{PU Vul} \citep{2012ApJ...750....5K}.

\begin{table}
\caption{Parameters of the WD during outburst (see text).}
\label{table:WDourburst}
\centering
\begin{tabular}{cccc}
\hline
\hline
Outburst phase & Year & L$_{\mathrm{WD}}$ [L$_\odot$] & T$_{\mathrm{WD}}$ [kK] \\
\hline
Maximum & 1980's & 8300--23000\\
Early nebular phase & 1993 & $3000\pm1000$ & $140\pm15$\\
Nebular phase & 2015 & $\sim$4200 & 291 \\
\hline
\end{tabular}

\end{table}

\section{Symbiotic nova classification}\label{sec:novaclass}

Based on the evolution of the WD parameters during the drop from maximum it is clear that the variability of \object{LMC~S154} is due to an outburst. However, a similar low-amplitude outburst could be explained by a Z~And-type outburst, and the SyN nature of the system cannot be confirmed by a light curve and WD parameters alone. 

We confirm the SyN nature of the object using spectroscopic features detected in the spectrum during nebular phase. \object{LMC~S154} shows a rich nebular spectrum, which is not observed in Z~And-type stars after outburst. In particular, the [\ion{N}{II}] forbidden lines are in strong emission in \object{LMC~S154}, while in the case of Z~And-type outbursts, forbidden lines with such low critical densities are never observed (see e.g., \citealt{2014MNRAS.444L..11M} and references therein).

Another strong argument in favor of the classical nova classification is provided by the evolution of the object in X-rays. In \object{LMC~S154,}  X-rays were detected only near the maximum of outburst and have not been detected since then \citep{1992ApJ...396..668R}. The opposite is expected in the case of Z-And-type outbursts, where the X-ray flux is lowest during outburst and highest during quiescence \citep{1997A&A...322..576G}.

\section{Conclusions and discussion}

While nova outbursts have been suggested for some extragalactic SySts, none have thus far  been proven. \textbf{This is perhaps mainly due to the fact that observations have never been carried out of all of the phases of an outburst in these systems.} Most notably, in the case of \object{Sanduleak's star,} outburst was suggested on the basis of monotonic fading  of the star \citep{2014MNRAS.438...35A}, but the beginning of the outburst was not observed in the history of the star, nor did the star reach quiescence. Moreover, an outburst was suggested in the case of M31SyS~J004322.50+413940.9 \citep{2014MNRAS.444..586M} but was never demonstrated.

In this work we study the possible evidence for the suggested nova outburst in the history of the extragalactic SySt, \object{LMC S154}. We gathered archival observations which confirmed that the variability of the system is due to a classical nova outburst. Results of our monitoring in the years 2005--2015 show that the outburst ended in 2009. This means that the star finally reached quiescence and all phases of an outburst were observed. Abundance of carbon in the photosphere of the RG is significantly higher than the abundance derived for the nebula, which confirms pollution of the circumbinary material by the nova ejecta. The luminosity and temperature changes of the WD are consistent with a nova outburst. The available evidence proves that \object{LMC~S154} is a bona fide SyN.

The archival data strongly suggest that there has been more than one outburst recorded in the history of \object{LMC S154}, which would make this system a member of a rare class of SyRNe. This makes this object the first SyRN in MCs and in general the third recurrent nova in the LMC \citep[see e.g.,][]{2016ApJS..222....9M}. While it is possible that \object{LMC~S154} showed only one nova outburst with one long maximum or more than one shorter maximum, as in the case of \object{RX Pup} \citep{1999MNRAS.305..190M}, the SyRNe classification is supported by the detection of  [\ion{O}{III}] lines between the two  observed maxima, in the 1940s and in the 1980s, and the nondetection in the 1980s. These lines are characteristic of the nebular phase of outburst, and once detected they last until the end of outburst, even if more than one maximum of the outburst is present (see e.g., \object{RX Pup}). This means that the maxima detected in the 1940s and the 1980s were associated with two different outbursts, and that each of them was followed by a nebular phase. Also in favor of the SyRN nature of \object{LMC S154} is the discovery of [\ion{Fe}{X}] and [\ion{Fe}{XI}] coronal lines. These lines are rare in SySts, and in SyNe they are observed only during the SSXS phase of nova outbursts on a massive WD. Another argument in favor of the recurrent nature of \object{LMC S154} is provided by the detection of X-ray radiation in the 1980s. The X-ray radiation from \object{LMC S154} most probably originated in a shock produced by an expanding blast wave, which is expected at the earliest phases of an outburst.

\object{LMC~S154} is the first classical nova to be discovered with a C-rich donor. While there have been no theoretical studies of nova outbursts with such a C-rich donor, it is clear that the abundance of carbon in the accreted material will have an influence on the evolution of the outburst. Together with the fact that \object{LMC S154} appears to suffer an outburst of a significantly longer timescale compared to other SyRNe (T~CrB, V745~Sco, V3890~Sgr and RS~Oph), this makes this object an interesting case for the study of nova outbursts.

\begin{acknowledgements}

KI has been financed by the Polish Ministry of Science and Higher Education Diamond Grant Programme via grant 0136/DIA/2014/43 and by the Foundation for Polish Science (FNP) within the START  program. This study has been partially founded by the  National  Science  Centre,  Poland,  grant OPUS  2017/27/B/ST9/01940. Polish participation in SALT is funded by grant No. MNiSW DIR/WK/2016/07.

We acknowledge with thanks the variable star observations from the AAVSO International Database contributed by observers worldwide and used in this research. This research has made use of data and/or software provided by the High Energy Astrophysics Science Archive Research Center (HEASARC), which is a service of the Astrophysics Science Division at NASA/GSFC and the High Energy Astrophysics Division of the Smithsonian Astrophysical Observatory. This research has made use of the VizieR catalogue access tool, CDS, Strasbourg, France. This publication makes use of VOSA, developed under the Spanish Virtual Observatory project supported from the Spanish MICINN through grant AyA2011-24052.
\end{acknowledgements}

\bibpunct{(}{)}{;}{a}{}{,} 
\bibliographystyle{aa} 
\bibliography{references} 

\begin{appendix}
\section{Photometric data}

\onecolumn
\longtab{
\begin{longtable}{ccccccccc}
\caption{\label{table:phot}Photometric observations of \object{LMC S154}.}\\
\hline
\hline
Date &  JD & m$_{\mathrm{pg}}$ / $B$ & $V$ & $R$ & $I$ / $Ic$ & $J$ & $K$ &  Ref.\\
 &  -2400000 & [mag] & [mag] & [mag] & [mag] & [mag] & [mag] &  \\
\hline
1975-11-28 & 42744 & 14.85  & & & & & & 1\\ 
1975-11-29 & 42745 & 14.83$\pm$0.4    & & & & & &  2\\
1976-01-04 & 42781 & 14.88$\pm$0.4    & & &  & & & 2\\ 
1982-05-07 & 45096 & 14.6       & & 12.5 &  & & & 3 \\
1982-11~~~~~ & & & & 12.95 &  & & & 4 \\
1982-11~~~~~ & & 14.72 & & 13.52 & 13.02 & & & 4 \\
1986-02-11 & 46472 & 14.35 & 13.90 & 13.28 &   & & & 5 \\
1987-03-28 & 46882 & 15.34 & 15.12 &      &  & & & 5 \\
1989-11-11 & 47841 & 15.14 & 15.30 & 13.65 &  & & & 5 \\
1993-11-22 & 49313 & 16.8& 15.7 & 14.2 &  & & & 6 \\
1997-03-02 & 50509 & & & & 14.25$\pm$0.03 & 12.18$\pm$0.06 & 9.64$\pm$0.08       & 7 \\
1998-10-25 & 51111 & & & & 13.69$\pm$0.04 & 11.5$\pm$0.06 & 9.13$\pm$0.10        & 7 \\
1998-11-23 & 51140 & & & & 13.74$\pm$0.03 & 11.72$\pm$0.06 & 9.40$\pm$0.06       & 7 \\
1998-12-02 & 51149 & & & &  & 11.71$\pm$0.03 & 9.38$\pm$0.02 & 8   \\
2004-11-01  &  53311  &   &  15.9$\pm$0.1  &   &    &   &   &  9  \\ 
2004-11-10  &  53320  &   &  16.0$\pm$0.1  &   &    &   &   &  9  \\ 
2004-12-12  &  53352  &   &  15.8$\pm$0.1  &   &    &   &   &  9  \\ 
2005-01-04  &  53375  &   &  15.75$\pm$0.1  &   &    &   &   &  9  \\ 
2005-01-29  &  53400  &   &  15.7$\pm$0.1  &   &    &   &   &  9  \\ 
2005-03-07  &  53437  &   &  15.8$\pm$0.1  &   &    &   &   &  9  \\ 
2005-03-31  &  53461  &   &  15.9$\pm$0.1  &   &    &   &   &  9  \\ 
2005-09-02  &  53615  &   &  15.8$\pm$0.1  &   &    &   &   &  9  \\ 
2005-12-30  &  53735  &   &  15.6$\pm$0.1  &   &    &   &   &  9  \\ 
2006-02-19  &  53786  &   &  15.7$\pm$0.1  &   &    &   &   &  9  \\ 
2006-04-16  &  53842  &   &  15.9$\pm$0.1  &   &    &   &   &  9  \\ 
2006-06-04  &  53891  &   &  15.75$\pm$0.1  &   &    &   &   &  9  \\ 
2007-02-20  &  54152  &   &  15.95$\pm$0.1  &   &    &   &   &  9  \\ 
2007-04-20  &  54211  &   &  16.18$\pm$0.1  &   &    &   &   &  9  \\ 
2007-05-11  &  54232  &   &  16.17$\pm$0.1  &   &    &   &   &  9  \\ 
2007-08-20  &  54332  &   &  16.19$\pm$0.1  &   &    &   &   &  9  \\ 
2007-11-09  &  54414  &   &  15.9$\pm$0.1  &   &    &   &   &  9  \\ 
2007-12-29  &  54464  &   &  16.0$\pm$0.1  &   &    &   &   &  9  \\ 
2008-02-08  &  54505  &   &  16.02$\pm$0.1  &   &    &   &   &  9  \\ 
2008-03-06  &  54532  &   &  16.1$\pm$0.1  &   &    &   &   &  9  \\ 
2008-04-15  &  54572  &   &  16.3$\pm$0.1  &   &    &   &   &  9  \\ 
2008-05-07  &  54594  &   &  16.8$\pm$0.1  &   &    &   &   &  9  \\ 
2008-06-04  &  54622  &   &  16.5$\pm$0.1  &   &    &   &   &  9  \\ 
2008-07-22  &  54669  &   &  16.5$\pm$0.1  &   &    &   &   &  9  \\ 
2008-09-05  &  54714  &   &  16.2$\pm$0.1  &   &    &   &   &  9  \\ 
2008-10-22  &  54761  &   &  16.4$\pm$0.1  &   &    &   &   &  9  \\ 
2008-12-06  &  54807  &   &  17.1$\pm$0.1  &   &    &   &   &  9  \\ 
2008-12-19  &  54820  &   &  16.3$\pm$0.1  &   &    &   &   &  9  \\ 
2009-01-20  &  54852  &   &  16.5$\pm$0.1  &   &    &   &   &  9  \\ 
2009-02-20  &  54883  &   &  17.2$\pm$0.1  &   &    &   &   &  9  \\ 
2009-03-24  &  54915  &   &  16.45$\pm$0.1  &   &    &   &   &  9  \\ 
2009-04-17  &  54939  &   &  16.9$\pm$0.1  &   &    &   &   &  9  \\ 
2009-05-17  &  54969  &   &  16.6$\pm$0.1  &   &    &   &   &  9  \\ 
2009-07-04  &  55016  &   &  16.5$\pm$0.1  &   &    &   &   &  9  \\ 
2009-07-24  &  55036  &   &  17.05$\pm$0.1  &   &    &   &   &  9  \\ 
2009-10-14  &  55118  &   &  17.3$\pm$0.1  &   &    &   &   &  9  \\ 
2009-12-13  &  55178  &   &  16.6$\pm$0.1  &   &    &   &   &  9  \\ 
2010-01-30  &  55227  &   &  16.55$\pm$0.1  &   &    &   &   &  9  \\ 
2010-04-08  &  55295  &   &  16.4$\pm$0.1  &   &    &   &   &  9  \\ 
2010-05-05  &  55322  &   &  17.1$\pm$0.1  &   &    &   &   &  9  \\ 
2010-06-11  &  55358  &   & 16.15$\pm$0.40   &  &  & & & 10 \\
2010-06-21  &  55368  &   &  16.8$\pm$0.1  &   &    &   &   &  9  \\ 
2010-08-02  &  55410  &   &  16.24$\pm$0.1  &   &    &   &   &  9  \\ 
2010-09-13  &  55452  &   &  16.52$\pm$0.1  &   &    &   &   &  9  \\ 
2010-09-19  &  55458  &   &  16.4$\pm$0.1  &   &    &   &   &  9  \\ 
2011-01-03  &  55565  &   &  17.0$\pm$0.1  &   &    &   &   &  9  \\ 
2011-01-08  &  55570  &   &  16.8$\pm$0.1  &   &    &   &   &  9  \\ 
2011-01-24  &  55586  &   &  16.9$\pm$0.1  &   &    &   &   &  9  \\ 
2011-02-06  &  55599  &   &  17.3$\pm$0.1  &   &    &   &   &  9  \\ 
2011-02-06  &  55599 &   & 16.20$\pm$0.40   &  &  & & & 10 \\ 
2011-02-07  &  55600  &   &  16.4$\pm$0.1  &   &    &   &   &  9  \\ 
2011-02-13  &  55605 &   & 16.25$\pm$0.65   &  &  & & & 10 \\ 
2011-02-16  &  55608  &   & 15.30$\pm$0.30  &   &  & & & 10 \\ 
2011-02-27  &  55620  &   &  16.7$\pm$0.1  &   &    &   &   &  9  \\ 
2011-03-28  &  55649  &   &  16.35$\pm$0.1  &   &    &   &   &  9  \\ 
2011-04-10  &  55662  &   &  16.1$\pm$0.1  &   &    &   &   &  9  \\ 
2011-05-09  &  55691  &   &  16.4$\pm$0.1  &   &    &   &   &  9  \\ 
2011-05-23  &  55705  &   &  16.3$\pm$0.1  &   &    &   &   &  9  \\ 
2012-05-31  &  56078  &   & 16.00$\pm$0.55  &   &  & & & 10 \\ 
2012-06-25  &  56103  &   & 15.70$\pm$0.60   &  &  & & & 10  \\ 
2011-07-02  &  55744  &   &  16.53$\pm$0.1  &   &    &   &   &  9  \\ 
2011-08-02  &  55775  &   &  16.63$\pm$0.1  &   &    &   &   &  9  \\ 
2011-09-08  &  55812  &   &  16.8$\pm$0.1  &   &    &   &   &  9  \\ 
2011-10-22  &  55857  &   &  16.55$\pm$0.1  &   &    &   &   &  9  \\ 
2011-11-19  &  55885  &   &  16.75$\pm$0.1  &   &    &   &   &  9  \\ 
2011-11-28  &  55894  &   &  16.92$\pm$0.1  &   &    &   &   &  9  \\ 
2011-12-20  &  55916  &   &  16.75$\pm$0.1  &   &    &   &   &  9  \\ 
2012-01-25  &  55952  &   &  16.5$\pm$0.1  &   &    &   &   &  9  \\ 
2012-02-13  &  55971  &   &  16.65$\pm$0.1  &   &    &   &   &  9  \\ 
2012-02-20  &  55978  &   &  16.35$\pm$0.1  &   &    &   &   &  9  \\ 
2012-03-19  &  56006  &   &  16.15$\pm$0.1  &   &    &   &   &  9  \\ 
2012-03-31  &  56018  &   &  16.35$\pm$0.1  &   &    &   &   &  9  \\ 
2012-04-16  &  56034  &   &  16.0$\pm$0.1  &   &    &   &   &  9  \\ 
2012-04-27  &  56045  &   &  16.5$\pm$0.1  &   &    &   &   &  9  \\ 
2012-05-07  &  56055  &   &  16.15$\pm$0.1  &   &    &   &   &  9  \\ 
2012-08-16  &  56155  &   &  16.4$\pm$0.1  &   &    &   &   &  9  \\ 
2012-10-10  &  56211  &   &  16.75$\pm$0.1  &   &    &   &   &  9  \\ 
2012-11-01  &  56232  &   &  16.55$\pm$0.1  &   &  13.98$\pm$0.1  &   &   &  9  \\ 
2012-11-08  &  56239  &   &  16.7$\pm$0.1  &   &  14.1$\pm$0.1  &   &   &  9  \\ 
2012-11-13  &  56245  &   &  16.6$\pm$0.1  &   &  14.08$\pm$0.1  &   &   &  9  \\ 
2012-11-25  &  56257  &   &  16.45$\pm$0.1  &   &  14.09$\pm$0.1  &   &   &  9  \\ 
2012-12-07  &  56268  &   &  16.6$\pm$0.1  &   &  14.2$\pm$0.1  &   &   &  9  \\ 
2012-12-17  &  56279  &   &  16.7$\pm$0.1  &   &  14.14$\pm$0.1  &   &   &  9  \\ 
2012-12-26  &  56287  &   &  16.45$\pm$0.1  &   &  14.21$\pm$0.1  &   &   &  9  \\ 
2013-01-02  &  56295  &   &  16.65$\pm$0.1  &   &  14.22$\pm$0.1  &   &   &  9  \\ 
2013-01-13  &  56306  &   &  16.65$\pm$0.1  &   &  14.19$\pm$0.1  &   &   &  9  \\ 
2013-01-23  &  56316  &   &  16.8$\pm$0.1  &   &  14.21$\pm$0.1  &   &   &  9  \\ 
2013-02-04  &  56328  &   &  16.7$\pm$0.1  &   &  14.28$\pm$0.1  &   &   &  9  \\ 
2013-02-20  &  56344  &   &  16.8$\pm$0.1  &   &  14.22$\pm$0.1  &   &   &  9  \\ 
2013-03-03  &  56355  &   &  16.75$\pm$0.1  &   &  14.2$\pm$0.1  &   &   &  9  \\ 
2013-03-20  &  56372  &   &  16.5$\pm$0.1  &   &  14.25$\pm$0.1  &   &   &  9  \\ 
2013-04-06  &  56389  &   &  16.6$\pm$0.1  &   &  14.2$\pm$0.1  &   &   &  9  \\ 
2013-05-05  &  56418  &   &  16.6$\pm$0.1  &   &  14.08$\pm$0.1  &   &   &  9  \\ 
2013-06-06  &  56449  &   &  16.35$\pm$0.1  &   &  14.19$\pm$0.1  &   &   &  9  \\ 
2013-06-06  &  56449  &   & 15.75$\pm$0.30   &  &  & & & 10  \\ 
2013-07-07  &  56480  &   &  16.6$\pm$0.1  &   &  14.4$\pm$0.1  &   &   &  9  \\ 
2013-07-30  &  56503  &   &  16.75$\pm$0.1  &   &  14.63$\pm$0.1  &   &   &  9  \\ 
2013-09-01  &  56536  &   &  16.8$\pm$0.1  &   &  14.95$\pm$0.1  &   &   &  9  \\ 
2013-09-13  &  56548  &   &  17.1$\pm$0.1  &   &  14.98$\pm$0.1  &   &   &  9  \\ 
2013-10-08  &  56573  &   &  16.85$\pm$0.1  &   &  15.13$\pm$0.1  &   &   &  9  \\ 
2013-10-28  &  56594  &   &  16.85$\pm$0.1  &   &  15.1$\pm$0.1  &   &   &  9  \\ 
2013-11-21  &  56618  &   &  16.8$\pm$0.1  &   &  15.22$\pm$0.1  &   &   &  9  \\ 
2013-12-05  &  56632  &   &  17.15$\pm$0.1  &   &  15.15$\pm$0.1  &   &   &  9  \\ 
2013-12-18  &  56645  &   &  17.25$\pm$0.1  &   &  15.12$\pm$0.1  &   &   &  9  \\ 
2013-12-26  &  56653  &   &  17.1$\pm$0.1  &   &  15.16$\pm$0.1  &   &   &  9  \\ 
2014-01-11  &  56669  &   &  17.3$\pm$0.1  &   &  15.32$\pm$0.1  &   &   &  9  \\ 
2014-01-31  &  56689  &   &  17.15$\pm$0.1  &   &  15.25$\pm$0.1  &   &   &  9  \\ 
2014-02-21  &  56710  &   &  17.25$\pm$0.1  &   &  15.15$\pm$0.1  &   &   &  9  \\ 
2014-03-05  &  56722  &   &  17.05$\pm$0.1  &   &  15.22$\pm$0.1  &   &   &  9  \\ 
2014-03-21  &  56738  &   &  17.0$\pm$0.1  &   &  15.1$\pm$0.1  &   &   &  9  \\ 
2014-03-27  &  56744  &   &  17.0$\pm$0.1  &   &  14.96$\pm$0.1  &   &   &  9  \\ 
2014-04-05  &  56753  &   &  16.9$\pm$0.1  &   &  14.97$\pm$0.1  &   &   &  9  \\ 
2014-04-16  &  56764  &   &  16.8$\pm$0.1  &   &  14.86$\pm$0.1  &   &   &  9  \\ 
2014-04-24  &  56772  &   &  16.9$\pm$0.1  &   &  14.85$\pm$0.1  &   &   &  9  \\ 
2014-05-01  &  56779  &   &  16.7$\pm$0.1  &   &  14.71$\pm$0.1  &   &   &  9  \\ 
2014-05-09  &  56787  &   &  16.8$\pm$0.1  &   &  14.74$\pm$0.1  &   &   &  9  \\ 
2014-05-18  &  56796  &   &  16.4$\pm$0.1  &   &  14.68$\pm$0.1  &   &   &  9  \\ 
2014-05-28  &  56806  &   &    &   &  14.51$\pm$0.1  &   &   &  9  \\ 
2014-05-30  &  56808  &   &  16.5$\pm$0.1  &   &  14.46$\pm$0.1  &   &   &  9  \\ 
2014-06-11  &  56820  &   &    &   &  14.56$\pm$0.1  &   &   &  9  \\ 
2014-06-17  &  56825  &   &  16.25$\pm$0.1  &   &  14.44$\pm$0.1  &   &   &  9  \\ 
2014-06-28  &  56836  &   &  16.32$\pm$0.1  &   &  14.54$\pm$0.1  &   &   &  9  \\ 
2014-07-08  &  56846  &   &  16.55$\pm$0.1  &   &  14.56$\pm$0.1  &   &   &  9  \\ 
2014-07-21  &  56859  &   &  16.25$\pm$0.1  &   &  14.46$\pm$0.1  &   &   &  9  \\ 
2014-08-16  &  56885  &   &  16.15$\pm$0.1  &   &  14.48$\pm$0.1  &   &   &  9  \\ 
2014-08-27  &  56896  &   &  15.92$\pm$0.1  &   &  14.28$\pm$0.1  &   &   &  9  \\ 
2014-09-13  &  56913  &   &  16.02$\pm$0.1  &   &  14.38$\pm$0.1  &   &   &  9  \\ 
2014-09-21  &  56921  &   &  16.18$\pm$0.1  &   &  14.36$\pm$0.1  &   &   &  9  \\ 
2014-10-05  &  56935  &   &  16.2$\pm$0.1  &   &  14.29$\pm$0.1  &   &   &  9  \\ 
2014-10-18  &  56948  &   &  16.05$\pm$0.1  &   &  14.25$\pm$0.1  &   &   &  9  \\ 
2014-11-14  &  56975  &   &  16.3$\pm$0.1  &   &  14.36$\pm$0.1  &   &   &  9  \\ 
2014-11-26  &  56987  &   &  16.65$\pm$0.1  &   &  14.33$\pm$0.1  &   &   &  9  \\ 
2014-12-15  &  57007  &   &  16.16$\pm$0.1  &   &  14.29$\pm$0.1  &   &   &  9  \\ 
2015-01-07  &  57030  &   &  16.25$\pm$0.1  &   &  14.33$\pm$0.1  &   &   &  9  \\ 
2015-01-09  &  57032  &   &  16.4$\pm$0.1  &   &  14.31$\pm$0.1  &   &   &  9  \\ 
2015-01-29  &  57051  &   &  16.35$\pm$0.1  &   &  14.33$\pm$0.1  &   &   &  9  \\ 
2015-02-16  &  57070  &   &  16.34$\pm$0.1  &   &  14.46$\pm$0.1  &   &   &  9  \\ 
2015-02-27  &  57081  &   &  16.47$\pm$0.1  &   &  14.47$\pm$0.1  &   &   &  9  \\ 
2015-03-18  &  57100  &   &  16.49$\pm$0.1  &   &  14.49$\pm$0.1  &   &   &  9  \\ 
2015-03-28  &  57110  &   &  16.52$\pm$0.1  &   &  14.55$\pm$0.1  &   &   &  9  \\ 
2015-04-07  &  57120  &   &  16.57$\pm$0.1  &   &  14.49$\pm$0.1  &   &   &  9  \\ 
2015-04-23  &  57136  &   &  16.62$\pm$0.1  &   &  14.56$\pm$0.1  &   &   &  9  \\ 
2015-05-06  &  57149  &   &  16.5$\pm$0.1  &   &  14.58$\pm$0.1  &   &   &  9  \\ 
2015-05-16  &  57159  &   &  16.5$\pm$0.1  &   &  14.56$\pm$0.1  &   &   &  9  \\ 
2015-05-25  &  57168  &   &  16.33$\pm$0.1  &   &  14.45$\pm$0.1  &   &   &  9  \\ 
2015-07-08 & 57211 & 16.46$\pm$0.07  & 16.31$\pm$0.06  &   & & & &  11  \\
2015-07-15  &  57218  &   &  16.2$\pm$0.1  &   &  14.37$\pm$0.1  &   &   &  9  \\ 
2015-07-29  &  57232  &   &  16.437$\pm$0.113  &   &   &   &   &  12  \\ 
2015-08-04  &  57239  &  16.842$\pm$0.120  & 16.383$\pm$0.063  &   &   &   &   &  12  \\ 
2015-08-10  &  57245  &  16.929$\pm$0.090  & 16.203$\pm$0.073  &   &   &   &   &  12  \\
2015-08-20  &  57255  &  17.733$\pm$0.183  & 16.185$\pm$0.073  &   &   &   &   &  12  \\ 
2015-08-22  &  57256  &   &  16.4$\pm$0.1  &   &  14.25$\pm$0.1  &   &   &  9  \\ 
2015-08-25  &  57260  &  17.083$\pm$0.098  & 16.104$\pm$0.060  &   &   &   &   &  12  \\ 
2015-09-06  &  57272  &  16.777$\pm$0.084  & 16.400$\pm$0.064  &   &   &   &   &  12  \\ 
2015-09-04  &  57269  &   &  16.06$\pm$0.1  &   &  14.2$\pm$0.1  &   &   &  9  \\ 
2015-09-13  &  57278  &   &  16.22$\pm$0.1  &   &  14.14$\pm$0.1  &   &   &  9  \\ 
2015-09-21  &  57286  &   &  16.27$\pm$0.1  &   &  14.37$\pm$0.1  &   &   &  9  \\ 
2015-10-06  &  57301  &   &  16.56$\pm$0.1  &   &  14.27$\pm$0.1  &   &   &  9  \\ 
2015-10-13  &  57308  &   &  16.51$\pm$0.1  &   &  14.32$\pm$0.1  &   &   &  9  \\ 
2015-10-28  &  57323  &   &  16.47$\pm$0.1  &   &  14.35$\pm$0.1  &   &   &  9  \\ 
2015-11-06  &  57333  &   &  16.5$\pm$0.1  &   &  14.42$\pm$0.1  &   &   &  9  \\ 
2015-11-09  &  57336  &  16.876$\pm$0.062  & 16.336$\pm$0.025  &   &   &   &   &  12  \\ 
2015-11-12  &  57339  &  16.761$\pm$0.039  &  16.276$\pm$0.026  &   &   &   &   &  12  \\ 
2015-11-12  &  57339  &   &  16.249$\pm$0.021  & 16.719$\pm$0.071  &   &   &   &  12  \\ 
2015-11-16  &  57343  &   &  16.258$\pm$0.028  &   &   &   &   &  12  \\ 
2015-11-16  &  57343  &   &  16.25$\pm$0.1  &   &  14.39$\pm$0.1  &   &   &  9  \\ 
2015-11-18  &  57344  &  16.711$\pm$0.027  & 16.368$\pm$0.033  &   &   &   &   &  12  \\ 
2015-11-22  &  57349  &  16.820$\pm$0.105  &   &   &   &   &   &  12  \\ 
2015-11-25  &  57352  &  16.764$\pm$0.086  &  16.195$\pm$0.032  &   &   &   &   &  12  \\ 
2015-11-27  &  57354  &  16.730$\pm$0.110  &  16.317$\pm$0.039 &   &   &   &   &  12  \\ 
2015-11-24  &  57350  &   &    &   &  14.52$\pm$0.1  &   &   &  9  \\ 
2015-12-03  &  57360  &   &  16.42$\pm$0.1  &   &  14.44$\pm$0.1  &   &   &  9  \\ 
2015-12-14  &  57371  &   &  16.61$\pm$0.1  &   &  14.49$\pm$0.1  &   &   &  9  \\ 
2015-12-17  &  57374  &   &  16.45$\pm$0.1  &   &  14.47$\pm$0.1  &   &   &  9  \\ 
2015-12-26  &  57383  &  16.821$\pm$0.066  &  16.307$\pm$0.043  &   &   &   &   &  12  \\ 
2015-12-27  &  57384  &   &  16.47$\pm$0.1  &   &  14.46$\pm$0.1  &   &   &  9  \\ 
2016-01-06  &  57393  &   &  16.8$\pm$0.1  &   &  14.43$\pm$0.1  &   &   &  9  \\ 
2016-01-08  &  57396  &   &  16.35$\pm$0.1  &   &  14.41$\pm$0.1  &   &   &  9  \\ 
2016-01-19  &  57407  &   &  16.33$\pm$0.1  &   &  14.4$\pm$0.1  &   &   &  9  \\ 
2016-01-27  &  57415  &   &  16.48$\pm$0.1  &   &  14.41$\pm$0.1  &   &   &  9  \\ 
2016-01-31  &  57419  &  16.960$\pm$0.033  & 16.216$\pm$0.028  &   &   &   &   &  12  \\ 
2016-01-31  &  57419  &   &  16.4$\pm$0.1  &   &  14.44$\pm$0.1  &   &   &  9  \\ 
2016-02-08  &  57427  &   &  16.35$\pm$0.1  &   &  14.38$\pm$0.1  &   &   &  9  \\ 
2016-02-09  &  57428  &   &  16.4$\pm$0.1  &   &  14.33$\pm$0.1  &   &   &  9  \\ 
2016-02-14  &  57433  &   &  16.25$\pm$0.1  &   &  14.39$\pm$0.1  &   &   &  9  \\ 
2016-02-16  &  57435  &   &  16.23$\pm$0.1  &   &  14.31$\pm$0.1  &   &   &  9  \\ 
2016-02-20  &  57439  &  16.745$\pm$0.074  &  16.313$\pm$0.037 &   &   &   &   &  12  \\ 
2016-02-23  &  57442  &   &  16.21$\pm$0.1  &   &  14.34$\pm$0.1  &   &   &  9  \\ 
2016-02-27  &  57446  &   &  16.39$\pm$0.1  &   &  14.31$\pm$0.1  &   &   &  9  \\ 
2016-03-04  &  57452  &   &  16.25$\pm$0.1  &   &  14.33$\pm$0.1  &   &   &  9  \\ 
2016-03-09  &  57457  &   &  16.35$\pm$0.1  &   &  14.37$\pm$0.1  &   &   &  9  \\ 
2016-03-11  &  57459  &   &  16.52$\pm$0.1  &   &  14.37$\pm$0.1  &   &   &  9  \\ 
2016-03-14  &  57462  &   &  16.35$\pm$0.1  &   &  14.43$\pm$0.1  &   &   &  9  \\ 
2016-03-16  &  57464  &   &  16.32$\pm$0.1  &   &  14.32$\pm$0.1  &   &   &  9  \\ 
2016-03-17  &  57465  &  16.721$\pm$0.038  &  16.203$\pm$0.025 &   &   &   &   &  12  \\ 
2016-03-22  &  57470  &  16.792$\pm$0.064  &  16.524$\pm$0.117  &   &   &   &   &  12  \\ 
2016-03-26  &  57474  &   &  16.27$\pm$0.1  &   &  14.39$\pm$0.1  &   &   &  9  \\ 
2016-03-28  &  57476  &   &  16.62$\pm$0.1  &   &  14.37$\pm$0.1  &   &   &  9  \\ 
2016-04-01  &  57480  &   &  16.77$\pm$0.1  &   &  14.49$\pm$0.1  &   &   &  9  \\ 
2016-04-03  &  57482  &   &  16.45$\pm$0.1  &   &  14.49$\pm$0.1  &   &   &  9  \\ 
2016-04-06  &  57485  &   &  16.5$\pm$0.1  &   &  14.5$\pm$0.1  &   &   &  9  \\ 
2016-04-07  &  57486  &   &  16.46$\pm$0.1  &   &  14.53$\pm$0.1  &   &   &  9  \\ 
2016-04-10  &  57489  &   &  16.41$\pm$0.1  &   &  14.59$\pm$0.1  &   &   &  9  \\ 
2016-04-14  &  57493  &   &  16.35$\pm$0.1  &   &  14.5$\pm$0.1  &   &   &  9  \\ 
2016-04-23  &  57502  &   &  16.3$\pm$0.1  &   &  14.75$\pm$0.1  &   &   &  9  \\ 
2016-04-28  &  57507  &   &  16.37$\pm$0.1  &   &  14.72$\pm$0.1  &   &   &  9  \\ 
2016-04-30  &  57509  &   &  16.92$\pm$0.1  &   &  14.7$\pm$0.1  &   &   &  9  \\ 
2016-05-04  &  57513  &   &  16.9$\pm$0.1  &   &  14.67$\pm$0.1  &   &   &  9  \\ 
2016-05-10  &  57519  &   &  16.72$\pm$0.1  &   &  14.64$\pm$0.1  &   &   &  9  \\ 
2016-05-14  &  57523  &   &  16.42$\pm$0.1  &   &  14.67$\pm$0.1  &   &   &  9  \\ 
2016-05-20  &  57529  &   &  16.7$\pm$0.1  &   &  14.73$\pm$0.1  &   &   &  9  \\ 
2016-05-25  &  57534  &   &  16.83$\pm$0.1  &   &  14.82$\pm$0.1  &   &   &  9  \\ 
2016-06-01  &  57541  &   &    &   &  14.88$\pm$0.1  &   &   &  9  \\ 
2016-06-05  &  57545  &   &  16.54$\pm$0.1  &   &  14.81$\pm$0.1  &   &   &  9  \\ 
2016-06-08  &  57548  &   &  16.86$\pm$0.1  &   &  14.75$\pm$0.1  &   &   &  9  \\ 
2016-06-16  &  57556  &   &  16.63$\pm$0.1  &   &  14.82$\pm$0.1  &   &   &  9  \\ 
2016-06-21  &  57561  &   &    &   &  14.9$\pm$0.1  &   &   &  9  \\ 
2016-06-28  &  57568  &   &    &   &  14.92$\pm$0.1  &   &   &  9  \\ 
2016-08-01  &  57601  &   &  16.75$\pm$0.1  &   &  14.83$\pm$0.1  &   &   &  9  \\ 
2016-08-11  &  57611  &   &  17.07$\pm$0.1  &   &  14.88$\pm$0.1  &   &   &  9  \\ 
2016-08-27  &  57627  &   &  16.6$\pm$0.1  &   &  14.85$\pm$0.1  &   &   &  9  \\ 
2016-09-12  &  57643  &   &  16.7$\pm$0.1  &   &  14.84$\pm$0.1  &   &   &  9  \\ 
2016-09-22  &  57654  &  17.010$\pm$0.042  &  16.578$\pm$0.032  &   &   &   &   &  12  \\ 
2016-09-24  &  57655  &   &  16.93$\pm$0.1  &   &  14.79$\pm$0.1  &   &   &  9  \\ 
2016-10-02  &  57664  &  16.602$\pm$0.084  &  16.305$\pm$0.068 &   &   &   &   &  12  \\ 
2016-10-11  &  57672  &   &  17.05$\pm$0.1  &   &  14.83$\pm$0.1  &   &   &  9  \\ 
2016-10-12  &  57674  &  17.089$\pm$0.080  &  17.313$\pm$0.270 &   &   &   &   &  12  \\ 
2016-10-13  &  57675  &   &  16.88$\pm$0.1  &   &  14.71$\pm$0.1  &   &   &  9  \\ 
2016-10-20  &  57682  &   &  16.5$\pm$0.1  &   &  14.64$\pm$0.1  &   &   &  9  \\ 
2016-10-23  &  57685  &   &  16.63$\pm$0.1  &   &  14.73$\pm$0.1  &   &   &  9  \\ 
2016-10-25  &  57687  &   &  16.66$\pm$0.1  &   &  14.65$\pm$0.1  &   &   &  9  \\ 
2016-10-30  &  57692  &   &  16.79$\pm$0.1  &   &  14.66$\pm$0.1  &   &   &  9  \\ 
2016-11-06  &  57699  &   &  16.72$\pm$0.1  &   &  14.7$\pm$0.1  &   &   &  9  \\ 
2016-11-18  &  57711  &   &  16.52$\pm$0.1  &   &  14.76$\pm$0.1  &   &   &  9  \\ 
2016-11-25  &  57718  &   &  16.71$\pm$0.1  &   &  14.7$\pm$0.1  &   &   &  9  \\ 
2016-11-30  &  57722  &   &  16.66$\pm$0.1  &   &  14.82$\pm$0.1  &   &   &  9  \\ 
2016-12-01  &  57724  &   &  16.62$\pm$0.1  &   &  14.76$\pm$0.1  &   &   &  9  \\ 
2016-12-08  &  57730  &   &  17.05$\pm$0.1  &   &  14.79$\pm$0.1  &   &   &  9  \\ 
2016-12-11  &  57734  &   &  16.58$\pm$0.1  &   &  14.82$\pm$0.1  &   &   &  9  \\ 
2016-12-18  &  57741  &   &  16.59$\pm$0.1  &   &  14.81$\pm$0.1  &   &   &  9  \\ 
2016-12-25  &  57748  &   &  16.57$\pm$0.1  &   &  14.86$\pm$0.1  &   &   &  9  \\ 
2016-12-29  &  57752  &   &  17.02$\pm$0.1  &   &  14.88$\pm$0.1  &   &   &  9  \\ 
2017-01-04  &  57758  &   &  16.71$\pm$0.1  &   &  14.95$\pm$0.1  &   &   &  9  \\ 
2017-01-10  &  57764  &   &  16.69$\pm$0.1  &   &    &   &   &  9  \\ 
2017-01-11  &  57765  &   &  16.52$\pm$0.1  &   &  14.89$\pm$0.1  &   &   &  9  \\ 
2017-01-17  &  57771  &   &  16.68$\pm$0.1  &   &  14.97$\pm$0.1  &   &   &  9  \\ 
2017-01-28  &  57782  &   &  16.71$\pm$0.1  &   &  15.0$\pm$0.1  &   &   &  9  \\ 
2017-02-03  &  57788  &   &  16.77$\pm$0.1  &   &  15.0$\pm$0.1  &   &   &  9  \\ 
2017-02-06  &  57791  &   &  16.54$\pm$0.1  &   &  14.97$\pm$0.1  &   &   &  9  \\ 
2017-02-15  &  57800  &   &  16.7$\pm$0.1  &   &  14.88$\pm$0.1  &   &   &  9  \\ 
2017-02-17  &  57802  &   &  16.78$\pm$0.1  &   &  14.92$\pm$0.1  &   &   &  9  \\ 
2017-02-21  &  57806  &   &  16.62$\pm$0.1  &   &  14.92$\pm$0.1  &   &   &  9  \\ 
2017-03-05  &  57818  &   &  16.64$\pm$0.1  &   &  14.76$\pm$0.1  &   &   &  9  \\ 
2017-03-06  &  57819  &   &  16.62$\pm$0.1  &   &  14.78$\pm$0.1  &   &   &  9  \\ 
2017-03-13  &  57826  &   &  16.59$\pm$0.1  &   &  14.82$\pm$0.1  &   &   &  9  \\ 
2017-03-16  &  57829  &   &  16.57$\pm$0.1  &   &  14.78$\pm$0.1  &   &   &  9  \\ 
2017-03-21  &  57834  &   &  16.95$\pm$0.1  &   &  14.64$\pm$0.1  &   &   &  9  \\ 
2017-03-25  &  57838  &   &  16.55$\pm$0.1  &   &  14.66$\pm$0.1  &   &   &  9  \\ 
2017-03-26  &  57839  &   &  16.54$\pm$0.1  &   &  14.73$\pm$0.1  &   &   &  9  \\ 
2017-03-29  &  57842  &   &  16.55$\pm$0.1  &   &  14.69$\pm$0.1  &   &   &  9  \\ 
2017-04-02  &  57846  &   &  16.69$\pm$0.1  &   &  14.74$\pm$0.1  &   &   &  9  \\ 
2017-04-09  &  57853  &   &  16.57$\pm$0.1  &   &  14.67$\pm$0.1  &   &   &  9  \\ 
2017-04-11  &  57855  &   &  16.51$\pm$0.1  &   &  14.61$\pm$0.1  &   &   &  9  \\ 
2017-04-15  &  57859  &   &  16.5$\pm$0.1  &   &  14.6$\pm$0.1  &   &   &  9  \\ 
2017-04-17  &  57861  &   &  16.53$\pm$0.1  &   &  14.61$\pm$0.1  &   &   &  9  \\ 
2017-04-21  &  57865  &   &  16.42$\pm$0.1  &   &  14.56$\pm$0.1  &   &   &  9  \\ 
2017-04-27  &  57871  &   &  16.54$\pm$0.1  &   &  14.66$\pm$0.1  &   &   &  9  \\ 
2017-05-02  &  57876  &   &  16.57$\pm$0.1  &   &  14.62$\pm$0.1  &   &   &  9  \\ 
2017-05-04  &  57878  &   &  16.61$\pm$0.1  &   &  14.62$\pm$0.1  &   &   &  9  \\ 
2017-05-09  &  57883  &   &  16.64$\pm$0.1  &   &  14.59$\pm$0.1  &   &   &  9  \\ 
2017-05-12  &  57886  &   &  16.83$\pm$0.1  &   &  14.56$\pm$0.1  &   &   &  9  \\ 
2017-05-19  &  57893  &   &  16.62$\pm$0.1  &   &  14.65$\pm$0.1  &   &   &  9  \\ 
2017-05-28  &  57902  &   &  16.74$\pm$0.1  &   &  14.65$\pm$0.1  &   &   &  9  \\ 
2017-06-02  &  57907  &   &    &   &  14.62$\pm$0.1  &   &   &  9  \\ 
2017-07-12  &  57946  &   &  16.9$\pm$0.1  &   &  14.7$\pm$0.1  &   &   &  9  \\ 
2017-07-18  &  57952  &   &    &   &  14.76$\pm$0.1  &   &   &  9  \\ 
2017-07-19  &  57953  &   &  17.1$\pm$0.1  &   &  14.68$\pm$0.1  &   &   &  9  \\ 
2017-07-24  &  57958  &   &    &   &  14.74$\pm$0.1  &   &   &  9  \\ 
2017-08-14  &  57979  &   &  16.57$\pm$0.1  &   &  14.7$\pm$0.1  &   &   &  9  \\ 
2017-08-23  &  57988  &   &  16.58$\pm$0.1  &   &  14.6$\pm$0.1  &   &   &  9  \\ 
2017-08-30  &  57996  &   &  16.72$\pm$0.1  &   &  14.82$\pm$0.1  &   &   &  9  \\ 
2017-09-02  &  57998  &   &  16.69$\pm$0.1  &   &  14.76$\pm$0.1  &   &   &  9  \\ 
2017-09-21  &  58017  &   &    &   &  14.53$\pm$0.1  &   &   &  9  \\ 
2017-09-24  &  58021  &   &    &   &  14.67$\pm$0.1  &   &   &  9  \\ 
2017-10-13  &  58040  &   &  16.85$\pm$0.1  &   &  14.68$\pm$0.1  &   &   &  9  \\ 
2017-10-23  &  58050  &   &  17.12$\pm$0.1  &   &  14.7$\pm$0.1  &   &   &  9  \\ 
2017-10-27  &  58054  &   &  16.8$\pm$0.1  &   &  14.76$\pm$0.1  &   &   &  9  \\ 
2017-11-06  &  58064  &   &  16.78$\pm$0.1  &   &  14.66$\pm$0.1  &   &   &  9  \\ 
2017-11-16  &  58074  &   &  16.83$\pm$0.1  &   &  14.56$\pm$0.1  &   &   &  9  \\ 
2017-11-24  &  58082  &   &  16.84$\pm$0.1  &   &  14.62$\pm$0.1  &   &   &  9  \\ 
2017-12-07  &  58095  &   &  16.76$\pm$0.1  &   &  14.54$\pm$0.1  &   &   &  9  \\ 
2017-12-11  &  58098  &   &    &   &  14.62$\pm$0.1  &   &   &  9  \\ 
2017-12-14  &  58102  &   &  16.92$\pm$0.1  &   &  14.53$\pm$0.1  &   &   &  9  \\ 
2017-12-19  &  58107  &   &  16.77$\pm$0.1  &   &  14.52$\pm$0.1  &   &   &  9  \\ 
2017-12-20  &  58108  &   &  16.94$\pm$0.1  &   &  14.59$\pm$0.1  &   &   &  9  \\ 
2017-12-25  &  58113  &   &  16.83$\pm$0.1  &   &  14.58$\pm$0.1  &   &   &  9  \\ 
2017-12-26  &  58114  &   &  16.73$\pm$0.1  &   &  14.56$\pm$0.1  &   &   &  9  \\ 
2017-12-28  &  58116  &   &  16.98$\pm$0.1  &   &  14.61$\pm$0.1  &   &   &  9  \\ 
2017-12-30  &  58118  &   &  16.8$\pm$0.1  &   &  14.52$\pm$0.1  &   &   &  9  \\ 
2018-01-01  &  58120  &   &  16.76$\pm$0.1  &   &  14.52$\pm$0.1  &   &   &  9  \\ 
2018-01-04  &  58123  &   &  16.75$\pm$0.1  &   &  14.44$\pm$0.1  &   &   &  9  \\ 
2018-01-06  &  58125  &   &  16.73$\pm$0.1  &   &  14.53$\pm$0.1  &   &   &  9  \\ 
2018-01-09  &  58128  &   &  16.82$\pm$0.1  &   &  14.49$\pm$0.1  &   &   &  9  \\ 
2018-01-11  &  58130  &   &  16.59$\pm$0.1  &   &  14.51$\pm$0.1  &   &   &  9  \\ 
2018-01-12  &  58131  &   &    &   &  14.53$\pm$0.1  &   &   &  9  \\ 
2018-01-13  &  58132  &   &  16.6$\pm$0.1  &   &  14.47$\pm$0.1  &   &   &  9  \\ 
2018-01-16  &  58135  &   &    &   &  14.47$\pm$0.1  &   &   &  9  \\ 
2018-01-17  &  58136  &   &    &   &  14.46$\pm$0.1  &   &   &  9  \\ 
2018-01-28  &  58147  &   &  16.71$\pm$0.1  &   &  14.4$\pm$0.1  &   &   &  9  \\ 
2018-02-01  &  58151  &   &  16.68$\pm$0.1  &   &  14.42$\pm$0.1  &   &   &  9  \\ 
2018-02-04  &  58154  &   &  16.65$\pm$0.1  &   &  14.38$\pm$0.1  &   &   &  9  \\ 
2018-02-05  &  58155  &   &    &   &  14.44$\pm$0.1  &   &   &  9  \\ 
2018-02-07  &  58157  &   &    &   &  14.4$\pm$0.1  &   &   &  9  \\ 
2018-02-09  &  58159  &   &    &   &  14.37$\pm$0.1  &   &   &  9  \\ 
2018-02-14  &  58164  &   &  16.75$\pm$0.1  &   &  14.32$\pm$0.1  &   &   &  9  \\ 
2018-02-19  &  58169  &   &  16.74$\pm$0.1  &   &  14.34$\pm$0.1  &   &   &  9  \\ 
2018-02-22  &  58172  &   &    &   &  14.33$\pm$0.1  &   &   &  9  \\ 
2018-02-25  &  58175  &   &  16.66$\pm$0.1  &   &  14.32$\pm$0.1  &   &   &  9  \\ 
2018-03-03  &  58181  &   &    &   &  14.22$\pm$0.1  &   &   &  9  \\ 
2018-03-11  &  58189  &   &  16.51$\pm$0.1  &   &  14.22$\pm$0.1  &   &   &  9  \\ 
2018-03-17  &  58195  &   &    &   &  14.25$\pm$0.1  &   &   &  9  \\ 
2018-03-20  &  58198  &   &  16.47$\pm$0.1  &   &  14.21$\pm$0.1  &   &   &  9  \\ 
2018-03-22  &  58200  &   &  16.42$\pm$0.1  &   &  14.2$\pm$0.1  &   &   &  9  \\ 
2018-03-23  &  58201  &   &    &   &  14.19$\pm$0.1  &   &   &  9  \\ 
2018-03-26  &  58204  &   &  16.44$\pm$0.1  &   &  14.24$\pm$0.1  &   &   &  9  \\ 
2018-04-01  &  58210  &   &  16.53$\pm$0.1  &   &  14.19$\pm$0.1  &   &   &  9  \\ 
2018-04-06  &  58215  &   &  16.46$\pm$0.1  &   &  14.19$\pm$0.1  &   &   &  9  \\ 
2018-04-11  &  58220  &   &  16.45$\pm$0.1  &   &  14.1$\pm$0.1  &   &   &  9  \\ 
2018-04-12  &  58221  &   &    &   &  14.16$\pm$0.1  &   &   &  9  \\ 
2018-04-15  &  58224  &   &  16.43$\pm$0.1  &   &  14.18$\pm$0.1  &   &   &  9  \\ 
2018-04-17  &  58226  &   &  16.55$\pm$0.1  &   &  14.15$\pm$0.1  &   &   &  9  \\ 
2018-04-20  &  58229  &   &  16.56$\pm$0.1  &   &  14.1$\pm$0.1  &   &   &  9  \\ 
2018-04-26  &  58235  &   &  16.35$\pm$0.1  &   &  14.12$\pm$0.1  &   &   &  9  \\ 
2018-05-03  &  58242  &   &  16.5$\pm$0.1  &   &  14.14$\pm$0.1  &   &   &  9  \\ 
2018-05-09  &  58248  &   &  16.41$\pm$0.1  &   &  14.11$\pm$0.1  &   &   &  9  \\ 

\hline\end{longtable}
\tablebib{
(1): SuperCOSMOS Sky Surveys \citep{2001MNRAS.326.1295H}; (2): The HST Guide Star Catalog, Version 1.2 \citep{2001AJ....121.1752M};  (3): The USNO-A2.0 Catalogue \citep{1992AJ....103..638M}; (4): The USNO-B1.0 Catalog \citep{2003AJ....125..984M}; (5): \citet{1992ApJ...396..668R};  (6): \citet{1996A&A...307..516M}; (7): DENIS Catalogue toward Magellanic Clouds \citep{2000AAS..144..235C};  (8): 2MASS All-Sky Catalog of Point Sources \citep{2006AJ....131.1163S}; (9): INTEGRAL OMC \citep{2001ESASP.459..375G}; (10): This work (Kleinkaroo Observatory); (11): This work (Du Pont); (12) Observations of Bart Staels from the American Association of Variable Star Observers (AAVSO). AAVSO UID=000-BCZ-119
}

}
\twocolumn

\begin{table}
\caption{Swift UV observations of \object{LMC S154}.}
\label{table:photswift}
\centering
\begin{tabular}{ccccccccc}
\hline\hline
Date &  Exposure [s]  & UVM2 [mag]\\
\hline
2014-05-17      &       383     &       16.94   $\pm$   0.07\\
2014-10-08      &       234     &       16.43   $\pm$   0.07\\
2014-10-09      &       454     &       16.46   $\pm$   0.05\\
2014-10-10      &       462     &       16.37   $\pm$   0.05\\
2014-10-25      &       164     &       16.60   $\pm$   0.09\\
2014-11-28      &       360     &       16.39   $\pm$   0.06\\
2014-12-05      &       410     &       16.35   $\pm$   0.05\\
2014-12-30      &       169     &       16.41   $\pm$   0.08\\
2015-01-01      &       132     &       16.42   $\pm$   0.09\\
2015-01-06      &       581     &       16.43   $\pm$   0.05\\
2015-01-19      &       479     &       16.42   $\pm$   0.05\\
2015-01-22      &       152     &       16.41   $\pm$   0.08\\
2015-01-24      &       613     &       16.32   $\pm$   0.04\\
2015-01-26      &       242     &       16.57   $\pm$   0.07\\
2015-01-27      &       190     &       16.33   $\pm$   0.07\\
2015-05-05      &       509     &       16.41   $\pm$   0.05\\
2015-05-07      &       449     &       16.36   $\pm$   0.05\\
2015-05-10      &       745     &       16.35   $\pm$   0.04\\
2015-05-12      &       494     &       16.45   $\pm$   0.05\\
2015-06-16      &       159     &       16.16   $\pm$   0.07\\
2015-06-27      &       845     &       16.22   $\pm$   0.04\\
2015-07-01      &       200     &       16.24   $\pm$   0.07\\
2015-10-08      &       325     &       16.18   $\pm$   0.06\\
2015-10-09      &       679     &       16.09   $\pm$   0.04\\
2015-10-11      &       352     &       16.03   $\pm$   0.05\\
2015-11-06      &       130     &       16.09   $\pm$   0.08\\
2015-11-08      &       130     &       16.09   $\pm$   0.08\\
2015-11-19      &       444     &       16.10   $\pm$   0.05\\
2015-12-04      &       773     &       16.10   $\pm$   0.04\\
2016-07-25      &       290     &       16.13   $\pm$   0.06\\
2016-07-26      &       795     &       16.19   $\pm$   0.04\\
2016-07-27      &       1094    &       16.21   $\pm$   0.04\\
2016-07-28      &       895     &       16.21   $\pm$   0.04\\
2016-07-29      &       1596    &       16.23   $\pm$   0.03\\
2016-07-31      &       203     &       16.18   $\pm$   0.07\\
\hline
\end{tabular}
\end{table}

\end{appendix}
\end{document}